\documentclass[12pt, letterpaper, oneside]{article}

\usepackage{graphicx}
\usepackage{epstopdf}
\usepackage[body={6.5in, 9in}, right=1in, top=1in]{geometry}
\usepackage{amssymb} 
\usepackage{amsmath} 
\usepackage{hyperref}
\usepackage{sidecap}

\newcommand\ee{\end{equation}}
\newcommand\be{\begin{equation}}
\newcommand\eea{\end{eqnarray}}
\newcommand\bea{\begin{eqnarray}}
\newcommand{\sfrac}[2]{{\textstyle\frac{#1}{#2}}}
\newcommand\di{\partial}

\begin{document}

\begin{center}
\Large{\textbf{The incompressible fluid revisited: \\ vortex-sound interactions }} \\[.8cm]
\large{Solomon Endlich and Alberto Nicolis}
\\[0.4cm]
\small{\textit{ Department of Physics and ISCAP, \\ 
Columbia University, New York, NY 10027, USA}}

\end{center}

\vspace{.2cm}

\begin{abstract}
We study the dynamics of a nearly incompressible fluid via (classical) effective field theory. In the kinematical regime corresponding to near incompressibility (small fluid velocities and accelerations), compressional modes are, by definition, difficult to excite, and can be dealt with perturbatively. We systematically outline the corresponding perturbative expansion, which can be thought of as an expansion in the ratio of fluid velocity and speed of sound. Our analysis is fully relativistic, but, if desired, the non-relativistic limit can be taken at any stage. We demonstrate the potential of our method by some sample calculations: the relativistic corrections to the emission and to the scattering of sound by vorticose (possibly turbulent) motions,
and, more interestingly, the vortex-vortex long range interaction  mediated by sound. These results are, to our knowledge, new. For the last one, we find a $1/r^3$ potential between localized vortex configurations (e.g., vortex rings), whose strength and detailed structure depend on a certain multipole moment of the vortices under consideration.
We also improve on the so-called vortex filament model, by providing a local field theory describing the dynamics of vortex-line systems and their interaction with sound, to all orders in perturbation theory.

\end{abstract}


\section{Introduction}

Hydrodynamics is  notoriously hard to solve. This is usually blamed on the non-linear nature of its equations of motion, and we will not quarrel with that.
It is instructive however to ponder what makes hydrodynamics stand out with respect to {\em other} non-linear field theories (general relativity, for instance), where  perturbative techniques exist that can be applied efficiently to a variety of physical situations.
Our position is that there are certain configurations---the vortices---that in general cannot be dealt with in perturbation theory. This is discussed at length in ref.~\cite{ENRW}, where the quantization of hydrodynamics was attempted, but the problem is there already at the classical level.

Consider, for instance, a non-relativistic incompressible fluid, where the only allowed configurations are vortices, i.e., divergence-free velocity fields. In that limit, the continuity equation is just the constraint $\vec \nabla \cdot \vec v= 0$, the only degree of freedom is  the vorticity field $\vec \omega = \vec \nabla \times \vec v$, and the Euler equation can be rewritten as
\be
 \frac{\di}{ \di t} \vec \omega = \vec \nabla \times \left( \vec v \times \vec \omega \right) \; .
\ee
If one neglects the non-linearity on the right hand side there is no time-evolution. The dynamics are completely dominated by non-linearities.

An equivalent viewpoint is the effective field theory one \cite{ENRW}: Vortices correspond to excitations that, because of an infinite-dimensional symmetry (volume-preserving diffeomorphisms), have vanishing energy in the limit of vanishing velocities, regardless of the excitations' spatial gradients.
It is then easy for the system to excite these zero-energy gradients, and to make them exit the regime of validity of perturbation theory.
Indeed, in field theory gradient energy counts as  quadratic potential energy in the Hamiltonian. In the absence of a quadratic potential, the dynamics are dominated by anharmonicities.

The situation is strikingly different for compressional modes (in a compressible fluid). At the equation of motion level they correspond to  perturbations $\delta \rho$ in the density field, obeying a wave equation plus non-linear corrections:
\be \label{compressible}
\Big(\frac{\di^2}{\di t^2}  - c_s^2 \nabla^2 \Big) \delta \rho + \dots = 0 \; .
\ee
Even neglecting the non-linearities, there are some non-trivial dynamics corresponding to the free propagation of sound waves. Then---for small $\delta \rho$---the non-linearities can be treated as small corrections to this. At the effective field theory level, compressional modes are standard gapless fields, which carry kinetic {\em and} gradient energy, and for which non-linear terms  correspond to standard perturbative interactions \cite{ENRW}.

These considerations indicate that there should exist physical situations in which compressional modes can be dealt with in perturbation theory, even though there might be an underlying vorticose `background' that cannot. A nearly incompressible fluid provides such a system; there, by definition, compressional modes are difficult to excite. 

It should be remembered  that near incompressibility is not an intrinsic property of certain fluids, but rather a kinematical regime that exists for {\em all} fluids: for fluid flows that are much slower than sound, any fluid behaves as nearly incompressible \cite{LL}, and vice versa, for fluid flows that are as fast as sound, any fluid is quite compressible. The analysis that follows thus applies to any fluid in the appropriate kinematical regime.

We will develop an effective field theory of sound interacting with vortices, and we will outline the systematics of the perturbative expansion that can be carried out close to the incompressible limit (sect.~\ref{coupling of vortex and sound}).\footnote{We will ignore dissipative effects due to viscosity and heat conduction throughout our paper. 
This is consistent at low frequencies and long wavelengths, because these effects correspond to higher order terms in the derivative expansion.
An initial attempt at incorporating dissipation in our effective field-theory language can be found in \cite{ENPW}.}
Some physical aspects of our analysis are not entirely new. For instance, the study of sound generation due to fluid flow (as opposed to time-dependent boundary conditions) was pioneered in the 1950's by Lighthill \cite{Lighthill1,Lighthill2}. Since then, its study has had a long history full of many contributions and contributors, the details of which can be found in the relatively recent texts \cite{Sound-Flow Interactions, Theory of vortex sound}. 

The basic idea of Lighthill is to identify the non-linear terms in \eqref{compressible} (the `dots') as the source of sound, and to try to solve that equation perturbatively.
In a sense our field theory is doing just that. However, by insisting on field theoretical ideas formulated in terms of a local action we can be
completely systematic in how we carry out such a perturbative expansion---simply because perturbation theory in quantum field theory has been exhaustively studied. We can organize each computation in terms of Feynman diagrams. By restricting to tree-level diagrams, we will be, in fact, just solving the classical equations of motion perturbatively. The Feynman diagram language is an extremely powerful organizational and computational tool to do that.

It should be noted that Lighthill's approach to eq.~\eqref{compressible} is the hydrodynamical analog of the so-called post-Netwonian expansion of general relativity (see e.g.~\cite{WeinbergGR}), which has been recently recast into an effective field theory language by Goldberger and Rothstein \cite{GR}. Our paper owes much to theirs.

Our first application of our tools will be to reproduce two preexisting results in the literature: the rate at which vorticose motions emit sound (sect.~\ref{Sound emitted by a vorticose source}), and the cross-section they have for scattering it (sect.~\ref{Scattering}). However, since our field theory is relativistic by construction, we will be able to provide the relativistic corrections to these results at no additional cost. To the best of our knowledge, these have never been computed before.

On top of providing powerful computational tools,  the effective field theory  also offers a novel viewpoint on the dynamics, which can lead to simple predictions of previously overlooked phenomena. For instance, since vortices interact with sound waves, and since these are gapless, from the field theoretical viewpoint it is {\em obvious}  that sound waves can mediate long-range interactions between vortices. These are in addition to the well known purely kinematical ``dragging" interactions, whereby the long-distance tail in each vortex's velocity profile  drags all the other vortices with it. We are not aware of any mention of our sound-mediated interactions in the literature. We will compute the leading contribution to the associated potential  (sect.~\ref{Potential between two vortices section}).

Interpreting the effects of such a potential on the vortices themselves will lead us to studying a somewhat different problem. The dynamics of vortices can be extremely counterintuitive. Even in the simplest case of vortex {\em lines}---which are the only possible vortices in superfluids, but exist in ordinary fluids as well---the equations of motion for the lines' positions are first order in time-derivatives. This means that concepts like that of `force' on a vortex  (which we would naively derive from our potential) do not really apply. We remedy this by constructing an action that describes the vortex lines dynamics in the purely incompressible limit (sect.~\ref{vortex lines}). This action includes the aforementioned ``dragging'' interactions as long-distance (that is, non-local) Lagrangian terms.
Then, our sound-mediated potential energy should just be interpreted as the first-order---in departures from the incompressible limit---correction to this action, whose consequences at the equation of motion level can be derived just by varying the action in the usual way.

Finally, bringing everything together (sect.~\ref{hydrophoton}), we rewrite the non-local dragging interactions as being mediated by an auxiliary local field, and we reintroduce the local couplings of vortices to sound, thus ending up with a convenient, {\em local } field theory describing the dynamics of vortex lines and their interactions---among themselves and with sound---to all orders in perturbation theory. We feel this to be  substantial technical as well as  conceptual improvement over the more standard ``vortex filament" model \cite{Donnelly}.

\section{Coupling of incompressible vortex flows to sound waves}
\label{coupling of vortex and sound}

We begin with the action for a relativistic perfect fluid. This can be written in two equivalent, alternative forms, depending on whether one treats, at any given time $t$, the fluid elements' physical (`Eulerian') coordinates $\vec x$  as functions of their comoving (`Lagrangian') coordinates $\phi^I$  \cite{JNPP},
\be
\vec x = \vec x (\phi^I, t) \; , \qquad I=1,2,3 \; , 
\ee
or vice versa \cite{Soper, DGNR, ENRW},
\be
\phi^I = \phi^I (\vec x ,t) \; .
\ee
Even though the former viewpoint is more intuitive from a mechanical viewpoint, the latter makes the construction of the action more transparent, since the $\phi^I$s transform simply as three scalar fields under the spacetime symmetries (Poincar\'e) acting on their argument $x^\mu= (t, \vec x)$. There is also a number of internal (i.e., coordinate independent) symmetries acting on the $\phi^I$ fields, most notably the volume-preserving diffeomorphisms \cite{DGNR, ENRW}
\be
\phi^I \to \xi^I (\phi^J) \; , \qquad \Big| \frac{\di \xi^I }{\di \phi^J}\Big| = 1 \; ,
\ee
corresponding physically to the absence of transverse stresses in a fluid.

Following standard effective field theory logic \cite{DGNR}, one finds that at long wavelengths and low frequencies, the action for the $\phi^I$  fields is given by
\be \label{action}
S = \int \! d^4 x \, F(b) \; , \qquad \text{with} \quad b \equiv \sqrt{\det \di_\mu \phi^I \di^\mu \phi^J } \; ,
\ee
and where $F$ is a generic function. A particular functional form of $F$ corresponds to a particular equation of state. For instance, various thermodynamic quantities such at the energy density, the pressure, the temperature, and the entropy density, can be expressed in terms of $F$ and its derivatives  \cite{DHNS}:
\be
\rho = -F \; , \qquad p = F - F_b b \; , \qquad T = - F_b \; , \qquad s = b \; , 
\ee
where the subscript $b$ denotes differentiation. This formalism can be straightforwardly extended to fluids at finite chemical potential for a conserved charge \cite{DHNS}. Notice that for an homogeneous, static fluid at equilibrium (at some given reference pressure), the physical coordinates coincide with the comoving ones,
\be \label{bkgrd}
\langle \phi^I \rangle_{\rm eq.} = x^I  \; .
\ee
This configuration has  $b=1$, which will be our reference value for $b$ for what follows.

Our goal here is to systematically incorporate compressional effects by expanding this Lagrangian around slow (with respect to the speed of sound) vorticose background fluid flows. At lowest order, this expansion was performed in \cite{ENRW}, but we reproduce it here for clarity. The vortex/compressional mode separation is made most clear by working first in the $\vec x( \phi,t)$ parameterization of the fluid, i.e. working in comoving coordinates. Additionally, we make the speed of light $c$ explicit in order to better keep track of relativistic effects. In these coordinates the action reads:
\be \label{comoving action}
S  =   - w_0 c^2 \int \! d^3 \phi dt \, \det J \,f\big( ( \det J^{-1}) \sqrt{1- v ^2/c^2} \big)  \; .
\ee
Where $J$ is the Jacobian matrix
\be \label{Jacobian}
J^i {}_j = \frac{\di x^i}{\di \phi^j}\; ,
\ee 
$\vec v$ is the fluid's velocity field,
\be
\vec v = \di_t {\vec x} ( \phi, t)|_{ \phi} \; ,
\ee
and $w_0 c^2$ is its enthalpy density at the reference point $b=1$:
\be \label{w0}
w_0 c^2 = (\rho+p)_{b=1} = - F_b(1) \; .
\ee
We are giving $w_0$ units of a mass density, and $f$ is a dimensionless function of its dimensionless argument,  defined by
\be \label{f(b)}
F(b) = - w_0 c^2 \, f(b) \; .
\ee
The above expression for the action is derived in Appendix \ref{expanding L} and was originally given in reference \cite{DGNR}. The form of equation (\ref{comoving action}) is convenient for taking the non-relativistic limit. However, we will find that keeping track of all relativistic effects is not particularly difficult and so we will do so.  

For incompressible flows we have
\be
\vec x( \phi,t) =\vec x_0 ( \phi,t) \; ,
\ee
where, at each moment in time, $\vec x_0 ( \phi,t)$ is a volume preserving diffeomorphism of the comoving coordinates $\vec \phi$,
\be
\det \left(J_0\right)=1 \; , \qquad J_0 \, ^i {}_j \equiv \det \left( \frac{\di x_0^i}{\di \phi^j}\right) \; .
\ee
This flow will only be a solution to the equations of motion, given by varying (\ref{comoving action}) with respect to $x^i (\phi,t)$, when the time-dependence of the flow vanishes. Consequently, for slow (relative to the speed of sound) vorticose flows, $x_0^i(\phi,t)$ is only an approximate solution with compressional corrections. So for these {\em nearly} incompressible flows we parametrize the fluid's configuration by
\be
\vec x( \phi,t)=\vec x_0( \phi,t)+ \vec \psi( \phi,t)
\ee
where $ \vec \psi (\phi,t)$ vanishes in the limit of negligible time-dependence of $\vec x_0( \phi,t)$, and is curl-free {\em with respect to $\vec x_0$}, in the sense that
\be
 \vec \psi = \vec \nabla_{x_0} \Psi (\vec x_0,t) \; ,
\ee
for some function $\Psi$.
It is easy to convince oneself that, at least to lowest order, this is the correct characterization of compressional modes. 
For instance the determinant of the Jacobian \eqref{Jacobian}, which characterizes the compression level of the fluid, at linear order in $\vec \psi$ is
\be
\det J \simeq 1 + \vec \nabla_{x_0 } \cdot \vec \psi \; ,
\ee
where we made use of the volume preserving property of $\vec x_0(\phi,t)$.
So, a curl component (wrt to $\vec x_0$) in $\vec \psi$ does not contribute to compression---only a gradient component does.
In the following we will denote derivatives wrt to $\vec x_0$ simply by $\vec \di_0$---not to be confused with the usual relativistic notation of denoting $\di_0$ as the partial derivative with respect to time, which will be denoted here simply by $\di_t$.

We can now expand (\ref{comoving action}) in terms of these small perturbations. To quadratic order in the compressional modes
we have
\be
\det J= 1+[\di_0 \psi]+\sfrac{1}{2}\left( [\di_0 \psi]^2-[\di_0\psi \di_0 \psi]\right)+ \dots \;,
\ee
where $[ \dots ]$ represents the trace \footnote{So $[\di \psi]= \di_i \psi^i$ and $[\di \psi \di \psi]= \di_j \psi^i \di_i \psi^j$ etc.}.
Meanwhile, the velocity field is given simply by
\be
\vec v= \vec v_0 +\di_t \vec \psi(\phi,t) \; ,
\ee 
where $\vec v_0$ is the divergence-free velocity field associated with $\vec x_0(\phi, t)$:
\be
\vec v _0= \di_t {\vec x_0} ( \phi, t)  \; , \qquad \vec \di_0 \cdot \vec v_0 = 0 \; .
\ee
Taylor-expanding and changing coordinates from the $\vec \phi$ to $\vec x_0$, and dropping total-derivative terms, the action can be written to lowest order as 
\be \label{lowest order}
S=S_0 +w_0  \int d^3 x_0 dt \; \Big\{ \sfrac{1}{2}\big(\sfrac{D}{Dt} \vec \psi \, \big)^2-\sfrac12{c_s^2}[\di_0 \psi]^2+ \vec v_0 \cdot \sfrac{D}{Dt} \vec \psi -\sfrac12 \sfrac{c_s^2}{c^2} \, v_0^2 [\di_0 \psi] + \dots \Big\}
\ee
where
\be \label{S_0}
S_0=w_0\int  d^3 \phi dt \, \sfrac{1}{2} \vec{v}_0 \, ^2+ \dots
\ee
has no $\psi$-dependence.
The $\dots$'s in $S_0$ are relativistic corrections, which are only powers of $\vec v_0 \, ^2$, $c$ and $c_s$. Since $S_0$ does not depend on $\vec \psi$, it should be thought of as the action for the incompressional flow parameterized by $\vec x_0(\vec \phi,t)$. To use $S_0$ in this sense, one should supplement it with a constraint that $\vec x_0$  be a volume-preserving diff of $\vec \phi$. For instance, it is straightforward to check that adding to $S_0$ the term
\be
\int \lambda(\phi, t) \big[ \det (\di x_0^i/\di \phi^j) - 1\big] \; ,
\ee
where $\lambda$ is a Lagrange multiplier, yields the correct Euler equation for incompressible fluids
\footnote{In the Euler equation, $\lambda$ plays the role of pressure.}.
However,  as we stressed in the Introduction, the incompressional dynamics are very difficult to solve for, whereas, at least for slow fluid flows, we can make substantial progress by systematically applying perturbation theory to the {\em compressional} degrees of freedom, for a given background  incompressible flow. So, in the following we will treat the volume-preserving evolution $\vec x_0(\phi, t)$
as given, and we will solve for the small compressional perturbations of such a background. In the quantum field theory/functional integral language, this is equivalent to performing the path integral---at tree level---over the $\vec \psi$ field, in the presence of given external sources $\vec x_0(\phi,t )$.

In \eqref{lowest order}, the $\frac{D}{Dt}$'s remind us that the partial time derivatives were taken at fixed {\em comoving} coordinates, i.e.~fixed $ \phi$. These are the usual convective derivatives of fluid dynamics. In the $x_0$ coordinates, they simply act as
\be
\sfrac{D}{Dt} g(x_0,t)=\di_t g(x_0,t)+(\vec v_0 \cdot \vec \di_0)g(x_0,t) \;.
\ee
This expression is exact, to all orders in $\vec \psi$:  it is a mathematical identity, which follows simply from the chain rule; 
the fact that $\vec x_0(\phi, t)$ only reproduces part of the full, physical flow, is irrelevant.
Finally, we have used that 
\be
f'(1) = 1 \; , \qquad f''(1) = c_s^2/c^2
\ee
($c_s$ is the sound speed), which follow, respectively, from the definitions \eqref{w0}, \eqref{f(b)}, and from a straightforward analysis of the perturbation spectrum about the background \eqref{bkgrd} \cite{DGNR, ENRW}.

And so, keeping the free quadratic terms in $\psi$ (those not coupled to $v_0$), and the linear coupling of $\psi$ to the background vortex velocity, we finally write the lowest order action containing the compressional modes as 
\begin{align}
\label{lowest order action}
\Delta S & = S_{\rm free} + S_{\rm int} \\
S_{\rm free} & =  w_0 \int d^3x_0 dt  \; \sfrac{1}{2}\Big( (\di_t \psi^i)^2-c_s^2 [\di_0 \psi]^2 \Big) \label{Sfree}\\
S_{\rm int} & =  w_0 \int d^3x_0 dt  \Big( v_0 ^i (\vec v_0 \cdot \vec \di_0)\psi_i-
\sfrac12 \sfrac{c_s^2}{c^2} \, v_0^2 [\di_0 \psi] + \dots \Big) \;. \label{Sint}
\end{align}
We have discarded a $\vec v_0 \cdot \di_t \vec \psi$ term after noting that, as $\vec \psi=\vec \di_0 \Psi$, we can place this spacial derivative on $\vec v_0$ by a simple integration by parts, and the term will vanish as $\vec \di_0 \cdot \vec v_0=0$.

This simple action is enough to generate the famous ``analogy'' of Lighthill \cite{Lighthill2}, with an additional relativistic correction. When we vary (\ref{lowest order action}) with respect to $\psi^i$ 
we recover Lighthill's limit of (\ref{compressible}). Of course, for the fluids Lighthill was implicitly considering the relativistic effects would be extremely subleading, however for ultra-relativistic fluids where $c_s^2$ can be as large as $c^2/3$ this correction term can become a 
leading-order effect.


\section{Sound emitted by a vorticose source}
\label{Sound emitted by a vorticose source}

 As an illustration of the utility of the effective field theory techniques, we reproduce Lighthill's result with additional relativistic correction terms. Instead of manipulating the lowest order equations of motion 
 we can work directly from the action given by (\ref{lowest order action}) utilizing the standard tools of quantum field theory to calculate physical observables. In particular, using the language of Feynman diagrams and restricting to tree level amplitudes allows us to isolate classical quantities \cite{Weinberg}. For comparison, a derivation of Lighthill's result using more standard techniques can also be found in Landau and Lifshitz's classic Fluid Mechanics textbook \cite{LL} \footnote{ It should be noted that there is a small typographical error in their final result.}.


Say that we have a  vorticose, possibly turbulent, configuration $\vec v(\vec x, t)$ with typical fluid flow of $v \ll c_s$ and some characteristic size $\ell$. The velocity field acts as a source for  sound waves. We can formally deal with this emission process via QFT tools, by computing the amplitude for emission of a single ``phonon" from our compact vorticose source. The linear (in $\vec \psi \,$)
terms of $S_{\rm int}$ in \eqref{lowest order action} contribute to the ``tadpole" diagram
\be \label{tadpole}
\begin{array}{c}\includegraphics[width=2.5cm]{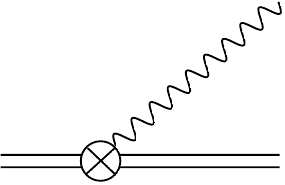} \end{array}=i \mathcal{M}(\vec p) \; ,
\ee
where $\mathcal{M}(\vec p)$ denotes the probability amplitude to emit a single phonon of momentum $\vec p$.
Given the probability amplitude one can calculate the emission rate with the standard formula:
\be\label{decayrate}
d\Gamma(\vec p)=\frac{1}{T}\frac{d^3p}{(2\pi)^3 2 c_s p }\left|\mathcal{M}(\vec p) \right|^2 \; ,
\ee
where $T\rightarrow \infty$  represents the total integration time,  and will drop out in time averaged quantities (which is what we are after). The standard relativistic formula has also been modified with $c\rightarrow c_s$ as the energy of a phonon with momentum $\vec p$ is simply $c_s p$
(see ref.~\cite{ENRW} and Appendix \ref{Scattering cross section from amplitude} for details).

Proceeding in the usual manner one finds:
\be
i \mathcal{M}(\vec p)= w_0^{1/2}\frac{p^i p^j}{p} \, \tilde{\kappa}^{*}_{ij}(c_s p, \vec p)\; 
\ee
where ${\kappa}_{ij}(\vec x, t)$ is the ``kinetic tensor''
\be \label{kappa}
{\kappa}_{ij}(\vec x, t) \equiv  v_{i} v_{j} - \sfrac12 \delta_{ij} \,  \sfrac{c_s^2 }{c^2} v^2 \; , 
\ee
and  $\widetilde{\kappa}_{ij}$ its Fourier transform,
\be
\tilde{\kappa}_{ij}(\omega, \vec p \, )=\int d^3 x dt \, {\kappa}_{ij}(\vec x, t) \; e^{i \omega t} e^{-i \vec p \cdot \vec x}\; .
\ee
Before inputting this amplitude into equation (\ref{decayrate}), we can make use of the multipole expansion: The typical frequency of the vorticose source is $\omega \sim v/\ell$. The emitted sound waves inherit such a frequency, and have therefore a typical wavelength $\lambda \sim c_s/ \omega \sim \ell \cdot (c_s/v) $, which is, in our approximation, much bigger than the vorticose source's size $\ell$. This means that we can treat the source as point-like, or the sound waves' wavelength as infinite. In the monopole approximation we have
\be
\tilde{\kappa}_{ij}(c_s p, \vec p \, ) \simeq \tilde{\kappa}_{ij}(c_s p, 0) \;.
\ee
%
%
In this approximation, integrating over the solid angle we have
\be \label{finaldecayrate}
d\Gamma( \omega) =\frac{\omega^3 \,  d\omega}{T} \frac{w_0}{60 \pi^2 c_s^5 }\Big[ |\tilde \kappa_{ii} |^2 +2 |\tilde \kappa_{ij}|^2  \Big] \;,
\ee
where the $\tilde \kappa$'s are evaluated at frequency $\omega$ and vanishing $\vec p$, and
we have used $\overline{n_i n_k}=\frac{1}{3}\delta_{ik}$ and $\overline{n_i n_k n_l n_m}=\frac{1}{15}(\delta_{ik}\delta_{lm}+\delta_{il}\delta_{km}+\delta_{im}\delta_{kl})$ (the `overline' denotes the average over the solid angle.)

The emission rate is not the most natural quantity to consider when calculating classical wave emission, a more typical one would be the power radiated. It is easy enough to augment (\ref{finaldecayrate}) in order to get this more standard quantity. If we have $\Gamma=\int d\Gamma $ we can write the power as simply $P=\int E \cdot d\Gamma $. In our case, the energy of a single phonon  is simply $\omega$ (we are working in $\hbar=1 $ units). Extending for convenience the range of integration to negative frequencies (and dividing by $2$), the total radiated power is
\be
P= \frac1{T} \frac{w_0}{120 \pi^2 c_s^5 } \int_{-\infty}^\infty d\omega\; \omega^4\Big[ |\tilde \kappa_{ii} |^2 +2 |\tilde \kappa_{ij}|^2  \Big] \;.
\ee
Noting that, for real $f(t)$ and $g(t)$,
\be
\int_{-\infty}^\infty \frac{d\omega}{2\pi} \, \tilde{f}(\omega) \, \tilde{g}^*(\omega) \, \omega^{2n}=\int_{-\infty}^\infty dt \frac{d^nf(t)}{dt^n} \,\frac{d^ng(t)}{dt^n} \; ,
\ee
we see that the inclusion of the $1/T$ factor, where $T\rightarrow\infty$, gives us time averages. Meanwhile, the restriction to vanishing $\vec p$ gives us volume integrals of the form
\be
{K}_{ij} (t) \equiv \int d^3 x \, \kappa_{ij}(\vec x, t) \; .
\ee
Putting everything together, for the radiated power we finally get
\be
P=\frac{w_0}{60 \pi c_s^5 } \Big[ \langle \ddot K_{ii} ^2 \rangle +2  \langle  \ddot K_{ij} {}^2 \rangle  \Big] \; .
\ee
where $\left< ... \right>$ denotes the time average, and the over-dots time derivatives.

To compare this result to the standard one---see e.g.~\cite{LL}---, we define the traceless and pure-trace quantities
\be
Q_{ij} (t) \equiv \int d^3 x \, \big( v_i v_j - \sfrac13 v^2 \delta_{ij} \big) \; , \qquad Q(t) \equiv  \int d^3 x  \,\sfrac13  v^2 \; , 
\ee
which we can use a basis for our tensor $K_{ij}$. In terms of these we get
\be
P=\frac{w_0}{\pi c_s^5 } \Big[ \sfrac{1}{4} \langle \ddot{Q} \, ^2\rangle \, \big(1-3\sfrac{c_s^2}{c^2}+\sfrac{9}{4}\sfrac{c_s^4}{c^4} \big) 
+\sfrac{1}{30}  \langle \ddot{Q}_{ij} {}^2\rangle \Big] \; .
\ee
This matches the result contained \cite{LL} with the addition of relativistic corrections. Note that these are much smaller than the leading terms  only if $c_s$ is much smaller than $c$, regardless of how small $v/c$ is. For relativistic equations of state with $c^2_s \sim 1/3 \, c^2$, the relativistic corrections are unsuppressed. The above expression is correct to lowest order in $v/c_s$---which is our small expansion parameter, associated with near incompressibility---but to {\em all} orders in $c_s/c$, and thus applies to those cases as well.

By going to higher orders in the diagrammatic expansion of eq.~\eqref{tadpole}, one could systematically compute higher order (in $v/c_s$) corrections to this result---for instance ``radiative" corrections, like was done in the case of gravitational wave emission by binary systems in \cite{GRoss}. The advantage of the EFT techniques for performing these higher-order computations is manifest: for instance, the UV divergences associated with the point-like approximation for the source can be handled in the standard and well understood ways of renormalization theory.


\section{Scattering sound waves off vorticose sources} \label{Scattering}

\begin{figure}[t]
\begin{center}
\includegraphics[width=0.8\textwidth]{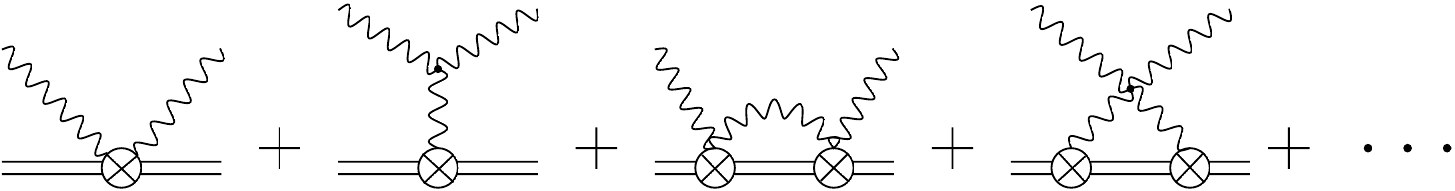}
\caption{\small \em Feynman diagrams contributing to the scattering of sound by a background fluid flow. The circled crosses denote $\psi$-$v_0$ interaction terms. There is  no propagator associated with the double lines, which just depict the external source $v_0$.
\label{fig: scattering}}
\end{center}
\end{figure}

If we include higher order terms in (\ref{Sint}) we can calculate other interesting quantities. For example, the scattering of sound waves by a nearly incompressible flow. As emphasized in \cite{LundRojas} this can be a powerful probe of the details of the fluid flow, e.g.~to address questions about  turbulence. The simplest scattering diagram will come from terms in the action quadratic in $\psi$---one $\psi$ for the incoming wave, one for the outgoing one---and coupled to $v_0$. In principle there are other diagrams contributing to the same physical process, depicted in fig.~ \ref{fig: scattering}.
However, it is easy to convince oneself by simple power counting that, for small $v_0/c_s$, the quadratic-in-$\psi$ vertex is the leading contribution: as derived in Appendix \ref{expanding L}, it involves a single power of $v_0$, whereas all the other diagrams start at order $v_0^2$. The `cross'  in the second diagram of the figure yields directly a $v_0^2$ factor, while the crosses  in the other diagrams yield a factor of $v_0$ each, but there are two such vertices per diagram.

Continuing the expansion that led to (\ref{Sint}), and which is summarized in detail in Appendix \ref{expanding L}, we obtain for this interaction term in the action:
\be
\label{simple quad action coupled to v}
S_{\rm int} \supset S_{\psi \, v^n} = w_0 \int d^3xdt \; \left\{ (\di_t \psi_i)(v \cdot \nabla \psi^i)-\frac{c_s^2 }{c^2}(v_i \di_t \psi^i)[\di \psi]  +{\cal O}(v^2) \right\}\; .
\ee
The second term is a relativistic correction, which is important if the speed of sound is not much smaller than that of light.
With an abuse of notation, we have stopped differentiating between $x$ and $x_0$, and between $v$ and $v_0$: all $x$'s and $v$'s should be thought of as being $x_0$'s and $v_0$'s.

When calculating the power emitted by a vorticose source in the previous section  we only made one assumption about the flow (in addition to its having a finite size): its flow velocity was very subsonic, with $v/c_s \ll1$. Now however, we have another dial we can control, that of the frequency of the injected sound waves. In particular, we take the same limit as \cite{LundRojas} and take the frequency of the incoming sound to be much larger than the natural frequency of the vorticose fluid flow: $\omega_\psi \gg \omega_v$. This is an experimentally convenient  limit, as it distinguishes the scattered radiation  from the naturally emitted one discussed in the previous section.

For given velocity profile $\vec v(\vec x, t) $ of the source, the $(\omega_1, \vec p_1) \to (\omega_2, \vec p_2)$ scattering amplitude associated with the interaction terms above is simply
\be
\label{one-to-one amplitude}
i {\cal M} =- i \Big\{(\hat p_1 \cdot \hat p_2)\left[ \omega_1 p_2^i+\omega_2 p_1^i\right]-\frac{c_s^2}{c^2}\left[ \omega_1  \hat p_1^i \, p_2 +\omega_2 \hat p_2^i \, p_1\right] \Big\} \tilde{v}^{i}(\Delta \omega, \Delta \vec p) \;, 
\ee
where $\Delta \omega$ and $\Delta \vec p$ are the energy and momentum transfers:
\be
\Delta \omega \equiv \omega_1- \omega_2 \; , \qquad \Delta \vec p \equiv \vec p_1 - \vec p_2 \; .
\ee
According to our assumptions, the typical frequency and wave-number of $\tilde v^i$ are
\be
\Delta \omega \sim \omega_v \ll \omega_{1} \; , \qquad \Delta p \sim \frac1\ell \sim \frac{\omega_v}{v} \; ,
\ee
where $\ell$ is the typical size of the source. We thus see that from the sound waves' viewpoint, the energy transfer is negligible, both in absolute terms and relative to the momentum transfer:
\be
\frac{\Delta \omega}{\omega_1} \ll 1 \; , \qquad \frac{\Delta \omega}{\omega_1}  \sim \frac{v}{c_s} \cdot \frac{\Delta p}{p_1} \ll \frac{\Delta p}{p_1}  \; .
\ee
To simplify the amplitude we can thus set
\be 
\omega_{1,2} \to \omega \; , \qquad |\vec p_{1,2}|\to \omega/c_s \; .
\ee
On the other hand, the momentum transfer---and thus the scattering angle---can be sizable, as long as $\omega_v/ \omega$ is not much smaller than $v/c_s \,$: 
\be
\frac{\Delta p}{p_1} \sim \frac{c_s}{v} \cdot  \frac{\omega_v}{\omega} \; .
\ee
We are thus left with
\be
i {\cal M} \simeq - i \, \frac {\omega^2}{c_s} \Big[(\hat p_1 \cdot \hat p_2)-\sfrac{c_s^2}{c^2}\Big] (\hat p_1+ \hat p_2) \cdot \tilde{\vec v}(\Delta \omega, \Delta \vec p) \; .
\ee
Inserting this  into the  standard formulae for the scattering cross section, which, for the convenience of the reader, are also quickly derived in Appendix \ref{Scattering cross section from amplitude},
we have
\be
\label{final result--sound scattering off source}
\frac{d\sigma}{d \Omega \,  d (\Delta \omega)}=\frac{1}{4c_s^6}\cdot\ \frac{\omega^4}{(2\pi)^3}\cdot\frac{1}{T}\left[ \left(\hat p_1 \cdot \hat p_2 \right)^2-\sfrac{2c_s^2}{c^2} \left(\hat p_1 \cdot \hat p_2 \right)+\sfrac{c_s^4}{c^4} \right] \big| (\hat p_1+\hat p_2)\cdot \tilde{\vec v} \big|^2\; ,
\ee
where $T \to \infty$ is a long observation time---long enough to observe the whole scattering process (the same $T$ should be used to define the Fourier transform $\tilde {\vec v}$).

If the source's velocity field is nearly stationary, $\tilde{\vec v} \sim (2\pi) \, \delta(\Delta \omega)$, or more generally if one does not have the experimental resolution necessary to detect the small frequency transfer  $\Delta \omega$, one can integrate in $\Delta \omega$ and end up with a time-averaged expression:
\be
\frac{d\sigma}{d \Omega}=\frac{1}{4c_s^6}\cdot\ \frac{\omega^4}{(2\pi)^2}\left[ \left(\hat p_1 \cdot \hat p_2 \right)^2-\sfrac{2c_s^2}{c^2} \left(\hat p_1 \cdot \hat p_2 \right)+\sfrac{c_s^4}{c^4} \right] 
(\hat p_1+\hat p_2)^i (\hat p_1+\hat p_2)^j
\big \langle  \tilde{v} ^i {}^* \tilde{v}^j \big \rangle \;  ,
\ee
where now  $\tilde{ v} = \tilde{v}(\Delta \vec p \, , t)$ denotes a { purely spatial} Fourier transform of the source's velocity field, and $\langle ... \rangle$ denotes a time-average.
Needless to say, the second and third terms in brackets are relativistic corrections, and---like in the previous section---they are exact to all orders in $c_s/c$.

The result expressed in (\ref{final result--sound scattering off source}) can be recast in a form that can be compared to previous results obtained by more traditional (and much more laborious) methods \cite{LundRojas}. However, the comparison takes some non-trivial translating and so it is contained in Appendix \ref{LundRojas matching}. When the smoke clears, we see that our answer disagrees with that given in \cite{LundRojas} by an overall factor of $2$. 
To test the self-consistency  of our computations, we check whether they obey the optical theorem. Since the optical theorem relates something quadratic in the scattering amplitude (the total  cross-section) to something linear in it (its  imaginary part in the forward limit), this is a non-trivial check of the overall normalization of our final result.
According to standard cutting rules, at this order in perturbation theory the optical theorem should relate the cross section we have just computed to the imaginary part of the the third diagram in fig.~\ref{fig: scattering}. As we show in 
 Appendix \ref{optical theorem check}, the check works out, which gives us confidence in our results.

\section{Sound-mediated long distance interactions between vortices}
\label{Potential between two vortices section}

In addition to the physical processes considered above, which have been previously treated  by perturbation theory performed at the level of the equations of motion (at least to lowest order for non-relativistic fluids), our formalism naturally motivates the calculation of the long-range interaction due to the exchange of compressional modes  between two physically separated vortex configurations.  The effective field theory approach {\em invites} such a question, and gives it a clear-cut qualitative answer: our vortices interact with compressional modes; these are gapless, and, as a consequence, can be exchanged between arbitrarily distant vortices; at some order in perturbation theory, this exchange will yield a non-trivial long range interaction between vortices. There is no question about it. For instance, this is how the classical $1/r$ Coulomb interaction arises in QED---via the tree-level exchange of virtual photons. On the other hand, such a question is not a  natural one to ask in the standard approach to fluid dynamics, and in fact, to our knowledge, these long-range interactions have never been postulated before---let alone computed.

Notice that there are other  long-range interactions between vortices, which survive even in the strict incompressible limit ($c_s\to \infty$), and are well understood in purely kinematical terms. As we will describe at some length below, each vortex typically carries a long distance ``tail'' in its  velocity profile, scaling  for instance as $1/r$ for infinitely long vortices. Then, if one is given several well-separated vortices,  each will be dragged by all the others' velocity tails. As will see, our sound-mediated interaction will  contribute  a small correction to this, suppressed by $(v/c_s)^2$, but by no extra powers of $r$.

In \cite{ENRW} we have initiated the analysis of our long-range interactions, by solving the linearized equation of motion for the sound modes and plugging the solution back into the action. Here we will use a more systematic technique---that of the effective action.\footnote{
An excellent pedagogical introduction to the effective action as utilized in a very similar context is given by \cite{LesHouches}.} 
Even though at the classical level using the effective action {\em is} equivalent to solving the (generally non-linear) equations of motion for some of the fields and plugging their solutions back into the action, and even though in the end we will restrict to a lowest order computation thus reproducing the result of \cite{ENRW}, we find it useful to set up a more systematic framework anyway, to pave the way for higher-order computations (which will appear elsewhere \cite{Solomon}) or for more general ones.

For our system, the sources will be given by  vortices characterized for simplicity by  the same typical length scale $l$ and the same typical velocity $v \ll c_s$, separated by a much bigger length scale $r \gg l$, so that they appear as point-like sources  to each other. 
These vortices interact via the exchange of compressional modes---virtual phonons, in our QFT-inspired language.
We can compute the potential energy due to this exchange, and standard thinking indicates that its gradient will give the ``force'' that the sources exert on each other. As we will see, this intuition is fundamentally {\em in}correct, but the potential energy is still a well-defined physical quantity worth computing, and we can worry about its dynamical implications  later.

\begin{figure}
\begin{center} 
\includegraphics[width=0.2\textwidth]{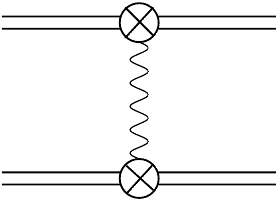}
\end{center}
\caption{\small \it Lowest order diagram contributing to the potential energy mediated by virtual sound between two vorticose sources.}
\label{2_vortex_interaction}
\end{figure}

The relevant quantity is the effective action $S_{\rm eff}$  one is left with after integrating out the compressional modes in the path integral. Schematically
\be
e^{i S_{\rm eff}[\vec x_0]} \equiv \int D \vec \psi \, e^{i S[\vec x_0, \vec \psi]} \; .
\ee
Since $\vec \psi$ is gapless,  $S_{\rm eff}$ will contain {\em non-local} interaction terms---precisely what we are after. As it turns out, the simple lowest order action given by (\ref{lowest order action}),
%
\be
S[\vec x_0, \vec \psi \, ] \simeq S_0 [\vec x_0 \, ]+ S_{\rm free} [\vec \psi \, ]+ S_{\rm int } [\vec x_0, \vec \psi \,] \; ,
\ee 
is all we need to perform this calculation to leading order.
$S_0$ is the lowest order action for $\vec x_0(\vec \phi,t )$, given in \eqref{S_0}. $S_{\rm free} $ is the quadratic action for $\vec \psi$, and $S_{\rm int}$ contains its linear couplings to $v_0$:
\be
S_{\rm int} = \int d^4 x \, \vec \psi \cdot \vec J \; , \qquad \vec J(x) \equiv w_0 \big(-\dot{\vec v}_0 -(v_0 \cdot  \nabla)\vec v_0 + \sfrac{1}{2} \sfrac{c_s^2}{c^2} \vec \nabla(\vec{v}_0 {} ^2) \big) \; .
\ee
Therefore, to lowest order we just have
\be
S_{eff}[\vec x_0]=S_0 - i\ln Z[\vec J \, ] \; , \qquad Z[\vec J \, ] \equiv \int \mathcal{D}\psi^i  e^{i S_{\rm free} + i \int d^4x \, \vec \psi \cdot \vec J+ \epsilon\, \text{terms}} \; ,
\ee
and we can compute the Gaussian functional integral via the standard completion of the square method. Up to an irrelevant constant we get  simply 
\be \label{lowest_order_Seff}
S_{\rm eff}=S_0+\frac{1}{2} \int d^4x d^4y \, J_i(x)\Delta^{ij}(x-y)J_j(y) \;,
\ee
where $\Delta^{ij}$ is the usual Feynman propagator defined by
\be \label{Feynman_prop}
 \Delta^{ij}(x-y) \equiv i \left< 0 \big|T\left\{ \psi^l(y)\psi^m(x) \right \} \big|0 \right> =\frac{1}{w_0}\int \frac{d^4k}{(2\pi)^4} \frac{\hat{k}^i \hat{k}^j e^{ik(x-y)}}{-k_0^2+c_s^2k^2-i\epsilon} \; .
\ee
Diagrammatically, this leading order correction to the effective action is given by the simple diagram denoted in fig. \ref{2_vortex_interaction}.

The compressional modes that mediate a long-distance interaction between slowly-moving sources are going to be (very) off-shell, i.e. $-k_0^2+c_s^2\vec{k}^2\ne 0$, and so we can take $\epsilon\rightarrow 0$ straight away. In particular, we expect the relevant frequencies and momenta to be
\footnote{We are assuming that $\dot v \sim v^2 /l$, so that the typical frequency of the sources is $v/l$. However, in many cases, like for instance those we will discuss below, one has nearly stationary, incompressible flows, in which case the typical frequencies are zero in first approximation, and the associated ``virtual'' compressional modes are maximally off-shell.}
\be
k^0 \sim \frac{v}{l}, \qquad \vec k \sim \frac{1}{r} \; ,
\ee 
and so if we take $v/c_s \ll l/r$ we can expand the denominator of the propagator as a power series in $(k^0/c_s k)$:
\be
\frac{1}{-k_0^2+c_s^2 {k}^2}=\frac{1}{c_s^2 {k}^2} \Big(1+\frac{k_0^2}{c_s^2\vec{k}^2}+... \Big)\; .
\ee
This is analogous to the low-energy expansion of a massive propagator in relativistic QFTs, and for us it means that our interactions, despite being genuinely non-local in space, can be expressed as a series of interaction terms that are at least local in time.

So, to lowest order in this expansion and isolating the interesting interaction term (that is, the cross term when we write the total vortex velocity as the sum of the two separate sources $\vec v =\vec v_1+ \vec v_2$) we have simply,
\begin{align} \label{ReSeff}
S_{\rm eff}= S_0 + \Delta S & = S_0+\frac{1}{w_0} \int d^4x d^4 y \frac{d^4 k}{(2\pi)^4}J_1^i(x) J_2^j(y) \frac{k^i k^j}{c_s^2 k^4}e^{ik(x-y)}\\
&=S_0 +w_0 \int_{x, y} \frac{d^4 k}{(2\pi)^4} \, \kappa_1^{ik} (x)\kappa_2^{jl} (y) \,  \frac{k^i k^k k^j k^l}{c_s^2 k^4}e^{ik(x-y)}\; ,
\end{align}
where we have freely integrated by parts and utilized the divergence free nature of $\vec v$, and the $\kappa$'s are the two sources' kinetic tensors, as defined in \eqref{kappa}.

Now, as these sources are supposed to be localized (that is, their velocity field falls off sufficiently fast away from their respective centers) we can multipole expand each source,  keep the lowest order term in $(l/r)$ (the monopole),  and perform the $k^0$ integral, which  generates a $(2 \pi) \delta (t_1-t_2)$. We may therefore write the integral of interest as:
\be
\Delta S \simeq \frac{w_0}{c_s^2} \int dt  \;  K_1^{ik}(t) K_2^{jl}(t)  \int  \frac{d^3 k}{(2\pi)^3} \; \frac{k^i k^j k^k k^l}{k^4} e^{i \vec k \cdot \vec r} \; ,
\label{lowest order potential}
\ee
where $\vec r$ is the vortices' relative position vector, and, like in sect.~\ref{Sound emitted by a vorticose source}, each $K$ is defined simply as the monopole moment of the corresponding $\kappa$, 
\be \label{monopole}
K^{ij} \equiv \int d^3 x \, \kappa^{ij} = \int d^3 x \, \big(v^{i} v^{j} - \sfrac12 \sfrac{c_s^2 }{c^2} \, \delta^{ij} \, v^2 \big) \; . 
\ee
The $\vec k$ integral is straightforward to perform by first rewriting the $k$'s at the numerator as gradients w.r.t.~$\vec r$, and then noticing that 
\be
\int\frac{d^3 k}{(2\pi)^3} \frac{1}{k^4} e^{i \vec k \cdot \vec r} = -\frac{1}{8\pi}\,  r + \mbox{IR-divergent, $\vec r$-independent pieces,}
\ee
as can be checked by introducing any IR-regulator, like e.g.~$1/k^4 \to 1/(k^2+m^2)^2$.

Recalling that the potential enters the action with an overall minus sign, we can finally write the effective vortex-vortex potential mediated by sound in a compact form as
\be \label{Delta V}
\Delta V =  \frac{w_0}{8 \pi c_s^2} \,   K_1^{ik} K_2^{jl} \, \di_i \di_j \di_k \di_l \, r \; .
\ee
The derivatives, once expanded,  yield
\be
\di_i \di_j \di_k \di_l \, r = \frac1{r^3} \big[ -\big(\delta_{ij} \delta_{kl} + \mbox{2 perms.}\big) -15 \cdot \hat r^i \hat r^j \hat r^k \hat r^l + 3 \big(\hat r^i \hat r^j \delta_{kl} + \mbox{5 perms.}\big)  \big] \; .
\ee

Notice that the  monopole moment \eqref{monopole} obeys positivity properties that forbids its vanishing as soon as one has {\em some} $v$---no matter how inventive one is in devising such a velocity profile. For instance, its trace is
\be
K^{ii} = \int d^3 x \, v^2 \big( 1- \sfrac32 \sfrac{c_s^2}{c^2}\big)  \; ,
\ee
which, using $c_s^2/c^2 \le 1/3$, is always bigger than $\int \frac12 v^2$---the kinetic energy per unit enthalpy.
As a result, the monopole-monopole interaction that we have computed will always be the most important long-distance interaction mediated by sound

For more general or  higher order computations it is probably more convenient to perform the multipole expansion directly at the level of the action, {\em before} integrating out $\psi$, along the lines of \cite{GR}. In this way, on top of organizing the perturbative expansion in a more systematic fashion, one is able to handle the UV divergences that unavoidably will show up at some order in perturbation theory using the standard tools of renormalization theory.
For instance, `self-energy' diagrams---diagrams in which a vortex exchanges a sound mode with 
itself---are UV divergent. Their divergent contribution to $S_{\rm eff}$ can be reabsorbed via renormalization into the  coefficients of terms that are already present in the (multipole-expanded) Lagrangian. We leave a systematic treatment of the multipole expansion for future work.

Eq.~\eqref{Delta V} is our final result for the long-distance, monopole-monopole interaction between localized vortex configurations. It is easy to estimate its size. The individual kinetic monopole moments are roughly $K^{ij}\sim v^2  l^3$, the derivatives structure acting on $r$ scales as $r^{-3}$, and so the effective potential scales like
\be
\label{delta Re V scaling}
\Delta V \sim  \frac{l^3}{r^3} \, \frac{v^2}{c_s^2} \cdot E_{\rm kin}\; ,
\ee
where $E_{\rm kin} \sim w_0 v^2 l^3$ it the typical kinetic energy of the vortex configuration. We will see in sect.~\ref{vortex lines} that the aforementioned kinematical dragging phenomenon can be modeled via a long-range potential energy that, for localized vortices, scales as $(l/r)^3 \cdot E_{\rm kin}$. Our effect is thus suppressed with respect to this one by an extra $(v/c_s)^2$ factor.


\subsection{Potential between two vortex rings}

\begin{SCfigure}
\includegraphics[height=5cm]{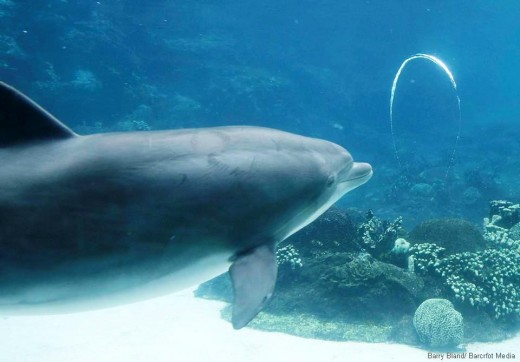}
\includegraphics[height=5cm]{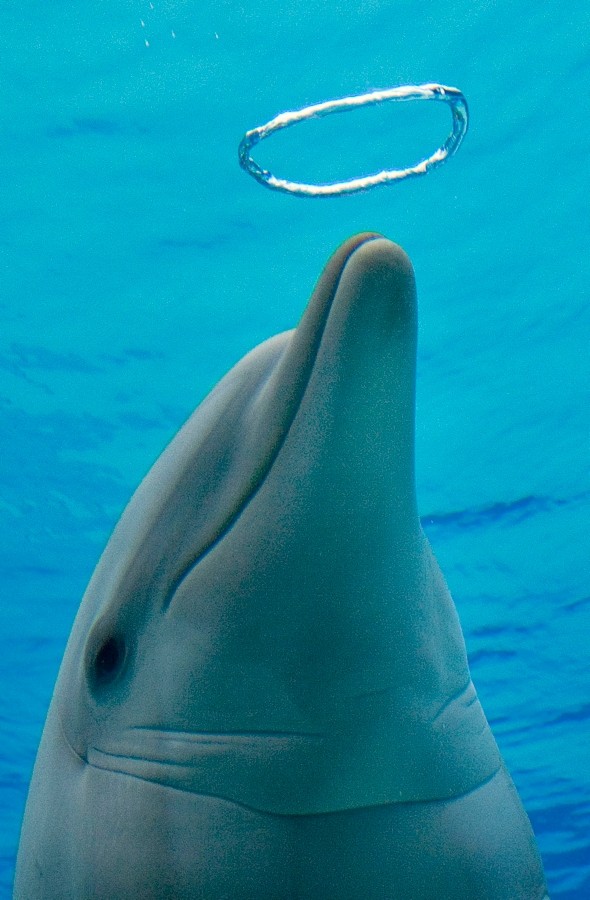}
\caption{\label{dolphin picture} \em \small  Dolphins can set up very clean vortex ring configurations, and play with them. They blow air into them, which, given the pressure gradients associated with the fluid flow, ``floats" to the core of the ring, thus serving beautifully as a tracer. Pictures taken from \cite{dolphin photo, dolphin photo 2}.}
\end{SCfigure}

In order to get a better physical sense of eq.~(\ref{Delta V}), we can compute it for the simple configuration of two interacting  circular vortex loops---the so-called vortex rings. Vortex rings are not only the simplest localized vortex-line configuration; they are also beautiful, fascinating objects of intense experimental interest and study, for humans \cite{Donnelly, WIrvine_DKleckner} as well as other species \cite{dolphins} (see fig.~\ref{dolphin picture}). For the following we will ignore the relativistic correction term in $K^{ij}$, but its inclusion is straightforward. We refer the reader to \cite{Donnelly} for an extensive review of the vortex rings' physical properties.


Say we have a vortex ring lying in some plane with normal vector $\hat{n}$. The radius of the ring is given by $R$, the circulation by $\Gamma$, and the vortex line thickness  by some cutoff $a$ (the so-called core radius). By symmetry $K^{ij}$ will take the form:
\be
K^{ij}=A \, \delta^{ij}+B \, \hat{n}^i \hat{n}^j \;.
\ee
The parameters $A$ and $B$ are given by the scalar integrals
\bea
A &=& \sfrac{1}{2} \int d^3 x \, \big( v^2 -(\vec v \cdot \hat n)^2  \big) \\
B&=& \sfrac{3}{2}\int d^3x \, \big( (\vec v \cdot \hat n)^2 -\sfrac{1}{3}v ^2 \big) 
\eea
For distances much larger than $R$ the velocity flow is given by that of a dipole,  that is, the velocity falls off like $1/r^3$, and so, as anticipated and desired, the main contribution to  $K^{ij}$ comes from the region near the vortex ring itself. However, at very close distances to the vortex line the same integral will diverge logarithmically (if the core radius $a$ is taken to zero). For $a \ll R$, we can thus get a good estimate of the integrals above by isolating the coefficient of $\log a$, which we can compute in the straight vortex line limit. Then, by dimensional analysis, $\log a$ has to be accompanied by $- \log R$. Geometric factors coming from precisely integrating over scales of order $R$ (which we can do, at least numerically) and from the details of how the UV cutoff is actually implemented (which is model-dependent), will show up as finite, order-one additions to this large universal $\log R/a$ contribution, and are therefore sub-leading.
Implementing this trick we find:
\be
K \simeq \sfrac14 \Gamma^2 R \,  \log (R/a) \left(\mathbb{I}+\hat{n} \otimes \hat{n} \right) \; .
\ee 
However for realistic vortex rings in normal fluids (as opposed to superfluids), $R$ is not that much bigger than $a$, and the  ``divergent'' log is so weak that one cannot neglect the allegedly sub-leading finite pieces. Therefore, in order to be completely general we can write
  \be
 K= \Gamma^2 R  \left( \alpha \, \mathbb{I}+ \beta \, \hat{n} \otimes \hat{n} \right) \; .
 \ee
 where $\alpha$ and  $\beta$ are order-one (or log) coefficients that in general depend on the exact core structure of the vortex rings.

Inserting this into (\ref{Delta V}) and assuming for simplicity the same core structure for the two rings, we can write the effective potential as 
\be
\label{V_eff vortex rings}
\Delta V= \frac{w_0}{8\pi c_s^2} \, (\Gamma^2_1R_1) \, (\Gamma^2_2  R_2)\cdot  \frac{1}{r ^3} \cdot f(\hat n_1, \hat n_2, \hat r) \; , 
\ee
where $f$ parameterizes the angular structure:
\begin{align}
f(\hat n_1, \hat n_2, \hat r) & =  -(\beta^2+4\alpha \beta) -2 \beta^2 (\hat{n}_1 \cdot \hat{n}_2)^2 -15 \beta^2 (\hat{n}_1\cdot \hat r)^2(\hat{n}_2\cdot \hat r)^2 \\
 &  +(6\alpha \beta+3\beta^2)\big[(\hat{n}_1\cdot \hat r)^2+(\hat{n}_2\cdot \hat r)^2 \big]+12\beta^2(\hat{n}_1\cdot \hat r)(\hat{n}_2\cdot \hat r)(\hat{n}_1 \cdot \hat{n}_2) \, . \nonumber
\end{align}
Notice that $f$ can be either positive or negative. For instance, if we take $\alpha=\beta=1$ for simplicity, then for a configuration with $\hat n_1= \hat n_2 = \hat r$ one gets $f=8$, while for $\hat n_1= \hat n_2 \perp \hat r$ one gets $f = -7$. That is, depending on the geometric configuration, this potential can be both ``attractive'' and ``repulsive''. As we will see in a moment, the quotes are in order, because these terms are extremely misleading for vortex dynamics.


\subsection{How (not) to interpret this $1/r^3$ potential}

For standard systems, one usually thinks of a potential like that given by (\ref{Delta V}) as a function of $\vec r$ that will generate a force in the usual way, $\vec F=- \frac{\di V}{\di \vec r}$. The problem  with this interpretation for us is that the notion of ``force'' does not really apply to vortex dynamics. To  see why this is the case, let us restrict ourselves to the study of vortex {lines}, which exist both in normal fluids and in superfluids, and of which the vortex rings  we just discussed  are  an example.

The way that we visualize a vortex line is by thinking of it as exactly that, a {\em line}, or more generally a curve. But really such a curve is a placeholder for a very special extended field configuration: the curve is the locus where the {\em vorticity} is nonzero, but the {\em velocity} field extends to large distances from the curve, typically in a $1/r$ fashion for very long lines. As a consequence, you cannot do to a vortex line the usual kinds of things  that you would to  a string-like  object in empty space, like, for instance, boost it. Doing so will change the boundary conditions at infinity of the velocity field, thus in effect boosting the whole fluid. This does not mean that vortex lines cannot move, however, it just simply means that for the velocity field to go to zero at infinity, the vortex line will move in only {\em one} particular way; there is no free initial condition for their velocity so that, at any given time, the local velocity of any vortex-line infinitesimal element is {\em completely} determined by the dynamics. For instance, isolated, perfect vortex rings always move at a constant speed determined by their radius and circulation (see for instance \cite{Donnelly}). Why do they do this? 

In the incompressible limit, as $\rho$ is no longer a dynamical variable, the hydrodynamical equations of motion are for $\vec v$ only and are given by the simplified continuity equation and the curl of the Euler equation:
\bea
\label{simplified continuity equation}
&&\vec \nabla \cdot \vec v=0\\
&& \frac{\di}{ \di t} \vec \omega = \vec \nabla \times \left( \vec v \times \vec \omega \right)\;  \label{vorticity eom}
\eea
where $\vec \omega =\vec \nabla \times \vec v$ is the vorticity field. The linearity of (\ref{simplified continuity equation}) implies that if we have multiple isolated vortex lines (lines where the vorticity is non-zero), then the velocity field outside these lines obeys linear superposition. Furthermore, Kelvin's theorem (which follows from \eqref{vorticity eom})
applied to infinitesimal comoving loops surrounding the vortex lines,
implies that each  line will move with the fluid flow generated by itself and every other vortex line \cite{somebody}.
Since the relation between $\vec \omega$ and $\vec v$ is formally identical to that between current density and magnetic field in magnetostatics, and since for vortex lines $\vec \omega$ is given by a sum of circulations times delta functions supported on the lines, with the circulations playing the role of total electric currents flowing along these lines,
the local velocity of the flow at $\vec x$ generated by all the vortex lines can be written in a Biot-Savart fashion as
\be
\label{first order eom}
\vec{v} (\vec x) = -\sum_{n}\frac{\Gamma_n}{4 \pi} \int_n \frac{(\vec x-\vec x \, ')}{|\vec x -\vec x \, '|^3} \times d\vec x \,' 
\ee
where $\Gamma_n$ is the circulation associated with the $n$-th vortex line
\footnote{There is a well-defined circulation $\Gamma_n = \int_{S_n} \vec \omega \cdot d\vec S$ for each line (the integral is taken over the cross-section of the ``line"), 
which is in particular constant along that line, because of  Stokes theorem, and constant in time, because of Kelvin's.}. 

As we can see, given that a vortex line element placed at $\vec x$ will move with this velocity, the vortex lines' positions obey {\em first order} equations of motion, in contrast to the usual second order equations of motions of any string-like object in empty space. This is the root of their peculiarity: with first-order equations of motion, there is no room for ``forces'', and the effective potential we have computed above must then be interpreted more carefully. Our strategy in the next section will be to derive a Lagrangian formulation that reproduces precisely these first-order equations motion. Then, our sound-mediated effective potential will just be a correction to such a Lagrangian, which will yield a corresponding correction to the equations of motion via the variational principle. 

The model that we briefly sketched above is called
the vortex-filament model and was pioneered in \cite{Schwarz 85} and \cite{Schwarz 88}. A modern review of such topics is discussed nicely in \cite{Donnelly}. In the next sections we will find a reformulation of the same model that we believe is likely to  lead to substantial progress in solving systems involving vortex lines.

Before we proceed, note that  the integral over the ``self-interacting'' line in \eqref{first order eom}, is going to diverge logarithmically as $\vec x \, ' \rightarrow \vec x$. Introducing a finite UV cutoff $a$, say the line's thickness, one gets that the leading contribution to the self-interacting integral is
\be
\int_{\rm self} \frac{(\vec x-\vec x \, ')}{|\vec x -\vec x \, '|^3} \times d\vec x \, '  \simeq \hat z \frac{1}{R}\log \big( R/a\big) + \dots \; ,
\ee
where $R$ is the line's local radius of curvature at $\vec x$, $\hat z$ is the direction  orthogonal to the plane defined by the local curvature, and the dots stand for subleading pieces, which come from the integral over regions that are far from $\vec x$, and which are finite for $a \to 0$.
This UV divergence tells us that there is a mild dependence of the velocity of vortex lines on the details of their core structure. Changing these details, or equivalently changing the definition of $a$, will only affect the subleading, finite pieces. As usual, the coefficient of the log divergence is universal. If one is in the large $\log({R/a})$ limit, the velocity is dominated by this universal piece. In the vortex filament model this approximation is called the local induction approximation, because the local velocity is mostly induced by the local curvature of the vortex line.

\section{An action for vortex-lines and their interactions}\label{vortex lines}

Given that this paper focuses on the Lagrangian formalism, it is natural to ask if there is an action that will generate (\ref{first order eom}) upon variation. As mentioned above, a byproduct of finding such an action is that it would provide us with a straightforward interpretation of our sound-mediated potential \eqref{Delta V}: this should simply be interpreted as a correction to the action, with obvious implications for the equations of motion.

Strictly speaking, the lowest order incompressible action \eqref{S_0}, $S_0= w_0 \int \frac{1}{2} \vec{v}_0 {} ^2$,  contains the dynamics we want to reproduce. 
However, we want to restrict the velocity field to configurations with a vorticity field  localized on lines, and we would like to use  the positions of the lines themselves as degrees of freedom. It is not trivial to rewrite $S_0$ as an action for these degrees of freedom.
So, for the moment we will just guess what the right action is, and we will comment later on what its relation to $S_0$ is.



Consider for the moment just one vortex line, parameterized by $\vec X(t, \lambda)$, where $\lambda$ is some parameter running along the line, over which we are going to integrate.
The action that we are going to write down should produce the equation of motion (\ref{first order eom}) upon variation w.r.t.~to $\vec X$.
Let's focus on the left hand side first. To get the velocity, which is just $\vec v = \di_t \vec X$, one needs a term in the action with {\em one} time derivative only. A nontrivial, natural candidate in 2D (for vortex {\em dots}) would be
\be
\int dt \,  \sfrac12 (X^1 \di_t  X^2 - X^2 \di_t X^1) = \int dt \,  \sfrac12 \epsilon^{ij} \, X^i \di_t X^j   \; .
\ee
Varying w.r.t.~to $X^1$ yields $\di_t X^2$, and varying w.r.t.~to $X^2$ yields $-\di_t X^1$. The antisymmetric structure is crucial in order not to end up with a total derivative. Then, a natural generalization to 3D seems to be
\be \label{XdXdX}
\int dt \, d \lambda \, \sfrac13 \epsilon^{ijk}X^i \di_t X^j \di_{\lambda} X^k  \; .
\ee
The vector $\di_{\lambda} \vec X$ is tangent to the curve, and so the combination $\epsilon^{ijk} \di_{\lambda} X^k$ behaves like the two-dimensional $\epsilon$-tensor for the plane locally orthogonal the curve, thus giving us back the 2D expression above. The only apparent downside is that when we vary w.r.t.~$\vec X$ to get the equations of motion, we now end up with an extra $\di_\lambda X^k$ multiplying (via a cross product) the velocity on the left hand side. 

Let's ignore this for the moment and let's move to the right hand side of \eqref{first order eom}. It contains a line integral over all vortex lines, including the one we are trying to understand the motion of. Then, the corresponding term in the action should  be represented by a double line integral, over all possible pairs of lines, including those pairs made up of the same line taken twice. 
Notice that the line element in \eqref{first order eom} can be written as $d \vec x \, ' \to d \lambda \, \di_\lambda \vec X$. Notice also that eq.~\eqref{first order eom} involves a cross-product. Given the extra cross product with $\di_\lambda \vec X$ we expect to get from \eqref{XdXdX}, and given that, schematically $\epsilon \epsilon \sim \delta$, we end up with the following inspired guess for the action of a system of vortex lines:
%
%
%
\be
\label{3D action for two vortex lines}
S = C  \int dt \bigg[  \sum_{n}  \int d\lambda \,  \sfrac{1}{3} \Gamma_n \epsilon^{ijk} \, X_n^i \, \di_t \, X_n^j \, \di_{\lambda} X_n^k 
+ \frac{1}{8\pi}  \sum_{n,m}  \int d \lambda d \lambda'  \Gamma_n \Gamma_m \frac{\di_{\lambda} \vec  X_n \cdot \di_{\lambda'} \vec X_m}{ |\vec{X}_n-\vec{X}_m  | }
\bigg] \; ,
\ee
where the sums run over all the vortex lines, $C$ is an overall---yet to be determined---constant (whose value does not affect the equations of motion), while the relative $1/8\pi$ is needed to get the correct equations of motion, as we will show in a moment. In this expression for the action it is understood that, in the double-integral term,  $\vec X_n$ is parameterized by $\lambda$ and $\vec X_m$ by $\lambda'$. As we stressed above, the double sum also includes  $n=m$, in which case the same line is traced out twice, independently by $\lambda$ and $\lambda'$.


Varying the curves in the usual way, $\vec{X}_n(\lambda,t) \longrightarrow \vec{X}_n(\lambda,t)+{\delta \vec X}_n(\lambda,t)$, and ignoring boundary terms, we get the equations of motion for the $n$-th line:
%
\begin{align}
\nonumber
\epsilon^{ijk} \, & \di_t X_n^j \, \di_{\lambda}X_n^k  \\
&= - \sum_{m}\frac{\Gamma_m}{4\pi} \int d\lambda' \left[ 
\frac{\big(\vec X_n- \vec X_m\big)^j \di_\lambda X_n^j \, \di_{\lambda'} X_m^i}{|\vec X_n- \vec X_m|^3}
-\frac{\big(\vec X_n- \vec X_m\big)^i \di_\lambda X_n^j \, \di_{\lambda'} X_m^j}{|\vec X_n- \vec X_m|^3} 
\right] \; ,
\end{align}
or, in vector notation (using  the identity $\vec a \times (\vec b \times \vec c) =
\vec b \, (\vec a \cdot \vec c) - \vec c \, (\vec a \cdot \vec b) $ on the right hand side):
\be \label{our kinetic vortex eom}
\di_t \vec X_n \times \di_{\lambda} \vec X_n =  - \left[ \sum_{m}\frac{\Gamma_m}{4\pi} \int d\lambda' \frac{(\vec X_n- \vec X_m\big) \times \di_{\lambda'} \vec X_m}{|\vec X_n- \vec X_m|^3} \right] \times  \di_{\lambda} \vec X_n  \; .
\ee

Now, it seems that our action (\ref{3D action for two vortex lines}) has not successfully reproduced the desired eom (\ref{first order eom}); there is an additional cross product with $\di_{\lambda} \vec X_n$  which we cannot simply remove from both sides of the equation. 
As a result, the equations of motion generated by our action are  a little less stringent then the ones generated by ``physical reasoning''. Should we be worried about this? 

As it turns out, the little extra freedom in (\ref{our kinetic vortex eom}) makes manifest a particular redundancy of how we describe the physical configurations of our system, which so far we have been reticent about. 
Given the spatial configuration of the vortex lines in our system at some particular time $t$, eq.~(\ref{our kinetic vortex eom}) will dictate the velocity of a particular point on the vortex line {\em only in the direction normal to that line}. That is, the component of the velocity {\em parallel} to the curve is undetermined. But as the fluid velocity flow is completely determined by the positions of the vortex lines, any motion that does not change the locations and shapes of the lines, such as motion {along} the lines, is totally unphysical.
In hindsight, it is obvious that this is related to a {\em reparameterization invariance} of the action \eqref{3D action for two vortex lines}: \be
\lambda \to \tilde \lambda (\lambda, t) \; .
\ee 
Both terms in the action are invariant, because the pieces that transform non-trivially are combined in structures that are manifestly invariant: at fixed $t$ one has
\be
d \lambda \, \di_\lambda \vec X_n = d \vec X_n \; , \qquad d \lambda \,  d \lambda' \, \di_\lambda \vec X_n \cdot \di_{\lambda'} \vec X_m =
d \vec X_n \cdot d \vec X_m\; .
\ee

This is a form of gauge invariance, which must be treated in the usual ways; for instance, in order to solve the equations of motion, one should first {\em choose a gauge}. The traditional form of the eom, (\ref{first order eom}), corresponds to a particular (and not so simple) 
gauge choice. 
Perhaps a more ``physical''  choice in our language would be to choose $\di_t \vec{X} \cdot \di_{\lambda}\vec{X}=0$, i.e~set the parallel component of the velocity to zero, for all time.
But depending on the problem at hand, other choices might be considerably more convenient. For instance, to study an infinite vortex line that is roughly aligned with a coordinate axis, say the $z$ axis, one can choose the gauge $\lambda = z \equiv X^3$, thus effectively
eliminating the $X^3$ degree of freedom from the problem.


\subsection{Fixing the overall coefficient}

While we have been able to reproduce the dynamics of the fluid system given the action (\ref{3D action for two vortex lines}), we still have  to determine the action's overall scaling. We can fix this constant $C$ by matching  a physical quantity for a simple physical configuration. As usual for effective field theory, the matching of Lagrangian coefficients can be done via an observable of an idealized, simple configuration, and the inferred values for the  coefficients can then be used for all other configurations as well. 

Consider a single vortex line aligned with the $z$-axis, with circulation $\Gamma$. Its velocity field is
\be
\vec v = \frac{\Gamma}{2\pi} \frac{1}{r} \,  \hat \varphi \; .
\ee
The original non-relativistic action for an incompressible fluid, eq.~\eqref{S_0}, implies that such a configuration has an energy per unit length
\be
\frac{d E }{d z} =w_0 \int \sfrac12 v_0^2 \, d^2x =  w_0 \frac{\Gamma^2}{4 \pi} \log{L/a}  \; , 
\ee
where $L$ is some IR cutoff (e.g.~the size of the container) and $a$ some UV cutoff (e.g.~the vortex-line thickness).\footnote{As before, as long as we are interested in the log-divergence only, we need not be specific about how these cutoffs are implemented, since the coefficient of the log is completely independent of these details---only additive finite pieces are sensitive to them.}

We can now compute the same physical quantity with our new action, eq.~(\ref{3D action for two vortex lines}). The first piece---the `kinetic' term---does not contribute to the Hamiltonian of the system, since it is linear in $\dot { X}$:
\be
E= H = \frac{\di L}{\di \dot { X}^i} \, \dot { X}^i - L \; .
\ee
We are thus left with the second piece only, which yields
\be
E = - C  \frac{\Gamma^2}{8\pi}   \int d z d z'   \frac{1}{ |z-z'  | } \; ,
\ee
corresponding to an energy per unit length
\be
\frac{d E }{d z} = - C \frac{\Gamma^2}{4 \pi} \log{L/a} \; .
\ee
Notice that, given the different geometric nature of the integrals involved, the cutoffs used here cannot have exactly the same meaning as those used above; the universality of log divergences allows us to be cavalier about this. Comparing the two expressions for the energy per unit length we get simply
\be
C = - w_0 \; .
\ee

\section{The hydrophoton} \label{hydrophoton}

Eq.~\eqref{3D action for two vortex lines} reproduces the correct dynamics and energetics of generic vortex line configurations, and is thus {\em the} correct action for them. Nevertheless, it has an obvious annoying feature: due to the double $\lambda$-integral in the second term, it is a {\em non-local} action. 
To develop an intuition about the (counter-intuitive) dynamics of the system, especially for perturbation theory questions (stability of solutions, properties and interactions of small perturbations, etc.), having a local action would be much more convenient.

Fortunately, given the $1/r$ nature of the non-locality in question, we know exactly how to fix the problem. We `integrate in' an auxiliary local field $\vec A(\vec x, t)$, coupled to the `currents' $\Gamma_n \, \di_\lambda \vec X_n$, which are localized on the vortex lines:
\begin{align} \label{local action}
S_{\rm local} = & \, w_0 \bigg[ -\sum_{n} \Gamma_n  \int dt  d\lambda     \,  \sfrac{1}{3} \epsilon^{ijk} \, X_n^i \, \di_t \, X_n^j \, \di_{\lambda} X_n^k  \\
 & + \int d^3 x dt \, \sfrac12 \big( \di_i A_j  \big)^2 -  \sum_{n}  \Gamma_n  \int dt d\lambda\,  \di_\lambda \vec X_n \cdot \vec A\big( \vec X_n, t \big) \bigg] \; . \nonumber
\end{align}
One can solve the equations of motion for $\vec A$ deriving from this action,
\be \label{maxwell}
\nabla^2 \vec A (\vec x, t) = -\sum_{n} \int  d\lambda\,  \Gamma_n \, \di_\lambda \vec X_n  \, \delta^3 \big( \vec x - \vec X_n (\lambda, t)\big) \; ,
\ee
and plug the solution back into action, thus getting back the non-local action \eqref{3D action for two vortex lines}. Since $\vec A$ appears at most quadratically in the action, this equivalence is {\em exact}, even at the quantum-mechanical level, which might be important for zero-temperature superfluids. 

In this new local picture, interactions can be described in standard field theoretical terms: the action at a distance in \eqref{3D action for two vortex lines} has been replaced by local interactions of each vortex-line with $\vec A$. Then, different vortex-lines, or different pieces of the same line, interact by `exchanging' $\vec A$.
It is obvious what $\vec A$ corresponds to in the magnetostatics analogy of the previous section: it is the vector potential in Coulomb gauge ($\vec \nabla \cdot \vec A = 0$). Indeed, given the (localized) current density
\be
\vec J (\vec x, t) = \sum_{n}   \Gamma_n  \int d\lambda\,  \, \di_\lambda \vec X_n  \, \delta^3 \big( \vec x - \vec X_n (\lambda, t)\big) \; ,
\ee
the $\int \vec J \cdot \vec A$ term in \eqref{local action} is precisely the standard interaction  between current and vector potential, and eq.~\eqref{maxwell} is precisely the equation of motion for the vector potential in Coulomb gauge, $\nabla^2 \vec A = \vec J$.
We dub $\vec A$ the `hydrophoton' field.
For more general vortex configurations in which the vorticity is not localized on lines, the analogy still works in detail---only, one should identify   the current density with the  vorticity field $\vec \omega$. Table \ref{dictionary} summarizes the dictionary between magnetostatics and our system.

\begin{table}[t]
\begin{center}
\begin{tabular}{r | l}
Magnetostatics & Incompressible Hydro \\
\hline
current $\vec J$ & vorticity $\vec \omega$ \\
magnetic field $\vec B$ & velocity field $\vec v$ \\
vector potential $\vec A$ & hydrophoton $\vec A$
\end{tabular}
\end{center}
\caption{\label{dictionary} \it\small Schematic dictionary between magnetostatics and our system of vortices in an incompressible fluid.}
\end{table}

In conclusion, the interaction between vortex lines in an incompressible fluid can be described {\em exactly} as magnetic interaction between currents. The only novelty is how the lines respond to these magnetic-type fields: given the peculiar structure of the lines' kinetic action in \eqref{local action} (the first line), there is no analog of the Lorentz force. Instead, as discussed in the previous section, the instantaneous velocity of a line-element {\em is} the local value of the `magnetic' field.

We can now assess very easily the importance of our sound-mediated potential \eqref{Delta V}, relative to the purely kinematical long-distance interactions which, in our new language, are mediated by $\vec A$.
For instance,  two vortex rings of radii $R_{1,2}$, circulations $\Gamma_{1,2}$, and orientations $\hat n_{1,2}$, at a large distance from each other have the same interaction potential energy as two magnetic dipoles,
\be
V_{\rm dip} = \frac{w_0}{r^3} \, \big[3 (\hat r \cdot \vec \mu_1) (\hat r \cdot \vec\mu_2) - \vec \mu_1 \cdot \vec \mu_2\big] \; ,
\ee
with dipole moments
\be
\vec \mu_n = \pi (\Gamma_n R_n^2) \, \hat n_n \; . 
\ee
We thus get
\be
V_{\rm dip}  \sim  \frac{w_0}{r^3}\, (\Gamma_1R^2_1) \, (\Gamma_2  R^2_2) \; ,
\ee
which is indeed a factor of $(c_s/v)^2$ bigger that the sound-mediated potential, eq.~\eqref{V_eff vortex rings}. Notice however that for very thin vortex cores, like e.g.~in superfluid vortex lines, eq~\eqref{V_eff vortex rings} gets enhanced by a factor of $(\log R/a)^2$, which can partially compensate the $(v/c_s)^2$ suppression factor.

The reader might be skeptical about the usefulness of our introducing the local field $\vec A$ to describe vortex interactions. In fact, aren't we supposed eventually to solve all the equations of motion? If we first solve the equation of motion for $\vec A$ and we plug the solution into the others, we are effectively reproducing the equations of motion deriving from \eqref{3D action for two vortex lines}. So, why bother introducing $\vec A$ in the first place? Although this is  technically a valid viewpoint---no information is added by introducing $\vec A$---such a statement is technically and conceptually equivalent to claiming that it is useless to introduce the local fields $\vec E$ and $\vec B$ (or $V$ and $\vec A$) in electrostatics and magnetostatics, since one can do everything at the level of the Coulomb force between charges and the magnetic force between currents.
A position that, in hindsight, few would defend. 
We believe that, like for electrostatics and magnetostatics, our local rewriting of vortex interactions will prove  valuable in the study of vortex line systems.

\subsection{Example: Kelvin waves}

To prove the usefulness of our approach, we now study via standard field-theoretical techniques the low-frequency spectrum of small perturbations of an infinite straight vortex-line. Let's align the (unperturbed) line with the $z$-axis. The perturbed fields are
\begin{align}
\vec X(\lambda,t) & = \vec X_0+\vec \pi (\lambda,t) \; , \qquad \vec X _0 = \hat z \lambda \\
\vec A (\vec x, t) & = \vec A_0+\vec \psi (\vec x, t) \; , \qquad \vec A _0 = -\hat z \frac{ \Gamma}{2 \pi} \log (r/a)
\end{align}
where $\vec X_0$ and $\vec A_0$ are the unperturbed configurations, $r = \sqrt{x^2+y^2} $ is the distance from the $z$-axis, and $a$ is a UV cutoff (the line thickness).
We can now expand $S_{\rm local}$ in powers of $\vec \pi$ and $\vec \psi$. The linear term in the fluctuations is, of course, going to vanish as $\vec X_0$ and $\vec A_0$ are solutions to the equations of motion. We are thus left with, at lowest order in the fluctuations, a quadratic action:
\bea
S& \simeq &- w_0 \bigg[ \int dt  d\lambda \, \sfrac{1}{2} \Gamma \epsilon_{ij} \pi_\perp^i \di_t \pi_\perp^j  - \int d^3 xdt  \, \sfrac{1}{2}\left(\di^i \psi^j \right)^2\
  \\
\nonumber
&&+  \Gamma \int dt d\lambda \, \Big( \vec \psi(0) \cdot  \di_\lambda \vec\pi  + \vec \nabla \psi^\parallel (0)  \cdot \vec  \pi  
+ \vec \nabla  A_0^\parallel (0) \cdot \vec  \pi \, \di_\lambda \pi^\parallel + \sfrac12 \di_i \di_j A_0^\parallel (0) \,  \pi^i \pi^j\Big)  \bigg]  \; ,
\eea
where `$\perp$' and `$\parallel$' are defined w.r.t.~the $z$-axis, and we have freely integrated by parts and thrown out boundary terms. 
In the second line, the  notation $\vec \psi(0)$ and  $\vec A_0(0)$ reminds us that these `bulk fields' and their derivatives are to be evaluated on the unperturbed line, at $x=y=0$
(the last three terms come from expanding their arguments in powers of $\vec \pi$).
One can thus see  that the second-to-last term vanishes as $\di^j A_0^\parallel$ is odd in either $x$ or $y$ (or simply vanishes for $j=3$).
Continuing further the decomposition of $\vec \psi$ and $\vec \pi$ into parallel and perpendicular components, and noticing that $\vec A_0$ only has derivatives in the orthogonal direction, we finally get
\bea
S& \simeq &- w_0 \bigg[ \int dt  d\lambda \, \sfrac{1}{2} \Gamma \epsilon_{ij} \pi_\perp^i \di_t \pi_\perp^j  -\int d^3 xdt  \, \sfrac{1}{2}\left(\di^i \psi_\perp^j \right)^2 +  \Gamma \int dt d\lambda \,  \vec \psi_\perp (0)\cdot  \di_\lambda \vec\pi _\perp \nonumber
  \\
&&-\int d^3 xdt   \sfrac{1}{2}\left(\di^i \psi_\parallel \right)^2 +   \Gamma \int dt d\lambda \, \Big(  \vec \nabla_\perp \psi_\parallel  (0) \cdot \vec \pi_\perp + \sfrac12 \di^\perp_i \di^\perp_j A_0^\parallel (0) \,  \pi_\perp^i \pi_\perp^j \Big)  \bigg]  \; . \label{full quadratic action for fluctuations}
\eea
%
%
Recall that we have not fixed the gauge yet. We would like to choose $\lambda = z = X^3$, that is,
\be
\pi_\parallel (\lambda, t) = 0 \; .
\ee 
Since $\pi^\parallel$ has completely disappeared from the action, we can perform this $\lambda \to z$ gauge choice directly at the level of the action without worrying about losing $\pi^\parallel$'s equation of motion. 

Varying \eqref{full quadratic action for fluctuations} with respect to $\psi_\perp$, $\psi_\parallel$, and $\pi_\perp$ we generate the three equations of motion:
\bea
 0&=& \nabla^2 \vec{\psi}_{\perp} +  \Gamma \delta^2(\vec{x}_\perp )  \, \di_z \vec \pi _\perp  \\
  0&=& \nabla^2 \psi_{\parallel} -  \Gamma \vec \nabla_\perp \delta^2(\vec{x}_\perp )  \cdot  \vec \pi _\perp  \\
    0&=&\di_t \vec \pi_{\perp} \times \hat z  - \di_z \vec \psi_\perp (0) + \vec \nabla_\perp \psi_\parallel (0)+\vec \nabla_\perp\di_i^\perp A_0^\parallel(0) \, \pi_\perp^i \; .
\eea
These equations can be recast into a slightly more convenient form by insisting on manifest translational invariance: The background configurations $\vec X_0(\lambda, t)$, $\vec A_0(\vec x,t )$ (spontaneously) break translations in the plane orthogonal to our line. As a results, these translations are non linearly linearized on the associated perturbations:
\be
\vec \pi _\perp \to \vec \pi_\perp + \vec \varepsilon_\perp \; , \qquad 
\psi_\parallel \to \psi_\parallel - \vec \varepsilon_\perp \cdot \vec \nabla_\perp A_0^\parallel \; ,
\ee
implying in particular---along the lines of Goldstone theorem---the absence of a gap at zero momentum. Although the action and equations of motion above are invariant under these transformations, the absence of a zero-momentum gap is not obvious at all, and will require non-trivial cancellations, since there are terms in which $\vec \pi_\perp$ appears without derivatives. It would be nicer to have a reformulation of these equations (and of the action), in which the broken translations act simply as a shift on $\vec \pi$ and nothing else. This is easily accomplished by redefining the $\psi_\parallel$ field as
\be
\psi _\parallel = \tilde \psi - \vec \pi_\perp \cdot \vec \nabla_\perp A_0^\parallel \; ,
\ee
so that, using $\nabla^2_\perp A_0^\parallel = - \Gamma \delta^2(\vec x_\perp)$, the eom become
\bea
 0&=& \nabla^2 \vec{\psi}_{\perp} +  \Gamma \delta^2(\vec{x}_\perp )  \, \di_z \vec \pi _\perp  \\
  0&=& \nabla^2 \tilde \psi -  \di_z^2 \vec \pi_\perp \cdot \vec \nabla_\perp A_0^\parallel \\
    0&=&\di_t \vec \pi_{\perp} \times \hat z  - \di_z \vec \psi_\perp (0) + \vec \nabla_\perp \tilde \psi (0) \; .
\eea
Now $\vec \pi_\perp$ always appears {\em with} derivatives, and translational invariance is realized simply as $\vec \pi _\perp \to \vec \pi_\perp + \vec \varepsilon_\perp$.
\footnote{The same change of variables at the level of the action, replaces the second line in \eqref{full quadratic action for fluctuations} with
\be
w_0 \int d^4 x \, \Big[ \sfrac{1}2 \big( \di_i \tilde \psi\big)^2 - \di_z \tilde \psi \,  \di_z \vec \pi_\perp \cdot \vec \nabla_\perp A_0^\parallel
+\sfrac12 \big( \di_z \vec \pi_\perp \cdot \vec \nabla_\perp A_0^\parallel\big) ^2 \Big] \; , 
\ee
which has manifest shift-invariance for $\vec \pi_\perp$. Varying the new action with respect to the fields one gets, after stratighforward manipulations, precisely the new equations of motion.
}

These equations describe waves propagating in the $z$ direction. To see this, let's go to  Fourier space for $z$ and $t$ (but not for $\vec x_\perp$), upon which the equations of motion become
\bea
0&=&  \big( \nabla_\perp^2 -k^2) \vec{\psi}_{\perp}+ \Gamma\, \delta^2(\vec{x}_\perp )  \, ik \,  \vec \pi_\perp  \\
0     &=& \big(\nabla_\perp^2 -k^2) \tilde \psi+k^2 \, \vec \pi_\perp \cdot \vec \nabla_\perp A_0^\parallel\\
      0&=&- i \omega\, \vec \pi_{\perp} \times \hat z  - i k \, \vec \psi_\perp (0) +\vec \nabla_\perp \tilde \psi (0) \; ,
\eea
where now $\vec \pi_\perp$ depends on $k$ and $\omega$, and $\vec \psi_\perp$ and $\tilde \psi$ depends on them as well as on $\vec x_\perp$.

The shape of the perturbed line is uniquely determined by $\vec \pi_\perp$. Therefore, as long as we are interested in just that, we only need the solutions for $\vec \psi_\perp$ and $\tilde\psi$ close to the line---see the third equation. Treating then $\nabla^2_\perp$ in the first two equations as being of order $1/r^2 \gg k^2$, close to the line we get
\begin{align}
&  \nabla_\perp^2  \vec{\psi}_{\perp}+  \Gamma \delta^2(\vec{x}_\perp )  \, ik \,  \vec \pi_\perp  \simeq 0 \qquad  \Rightarrow \qquad
\vec{\psi}_{\perp} \simeq - \frac{ \Gamma}{2 \pi } \, i k  \,  \log(kr) \, \vec { \pi}_\perp \\
 & \nabla_\perp^2 \tilde \psi - \frac{\Gamma}{2\pi} k^2 \, \vec \pi_\perp \cdot \frac{\vec r_\perp}{r^2} \simeq 0
 \qquad  \Rightarrow \qquad \tilde \psi \simeq \frac{\Gamma}{4\pi} k^2 \,  \log(kr) \,  \vec r_\perp \cdot \vec \pi_\perp 
\end{align}
(The exact solutions at all distances involve a Bessel function of the second kind, $K_0(kr)$, and have the precisely these asympthotics for $kr \ll 1$.) 
Plugging these solutions into the third equation, we finally get a wave equation for $\vec \pi_\perp$ only:
%
%
%
%
\be
-i \omega \, \big(\vec { \pi}_{\perp} \times \hat z \big)  +   \frac{\Gamma}{4\pi}k^2 \log(1/ka) \, \vec { \pi}_\perp=0 \; ,
\ee 
where we have stopped the limit $r \to 0$ at some UV cutoff $a$, on which, like before, the log weakly depends, and 
 we have ignored a piece proportional to $ r_\perp^i \di^j_\perp \log(kr) \big|_a$, which is finite for $a \to 0$, and can thus be thought of as redefining the UV cutoff $a$ inside the  log. 
Notice that we have expressed the log so that it is positive, since $ka \ll 1$.

Such a wave equation describes  two transverse, circularly polarized modes of oscillation, with dispersion relations
\be
\omega=\pm\frac{\Gamma}{4\pi}k^2 \log( 1/ka) \; , 
\ee  
and  polarizations $\frac1{\sqrt 2}(1, \mp i)$.
Note that both these modes circulate (at a fixed height along the vortex line) in the opposite direction of the flow circulation. 
These are the famous Kelvin waves, and our results match their qualitative and quantitative properties as derived via 
Victorian methods \cite{Lord Kelvin}.

Despite the (weak) dependence on the UV cutoff, the dispersion relation above has predictive content. We can rewrite it in an `RG-invariant' fashion by comparing the frequencies associated with two different momenta:
\be
\frac{\omega_1}{k_1^2} - \frac{\omega_2}{k_2^2} =  - \frac{\Gamma}{4\pi} \log (k_1/k_2) \; ,
\ee
or equivalently
\be
\frac{\di (\omega/k^2)}{\di \log k } =  -  \frac{\Gamma}{4\pi} \; .
\ee
%
Such  relations are completely independent of the UV details of the vortex line, and are thus a robust prediction of the low-energy effective field theory. For instance, one can use them to {\em measure} indirectly  the circulation $\Gamma$.

\subsection{Reintroducing sound}
It is now time to go back to the first part of our paper, and look again at how compressional modes couple to incompressional fluid flows.
Consider first the lowest-order coupling \eqref{Sint}, and, to be consistent with this section and the previous one, let's neglect the second term, which is a relativistic correction. In place of $\vec v_0$, we could plug in the integral expression \eqref{first order eom}, which gives the incompressional velocity field as a function of the vortex lines' geometry. Clearly, the resulting Lagrangian term would correctly describe the interaction between vortex lines and sound, and can obviously be extended to all orders in $(v/c_s)^2$, following the expansion of Appendix \ref{expanding L}. The downside of this approach is once again the non-local nature of the resulting action: for each power of $v_0$, we end up with a line integral over all vortex lines.

The hydrophoton provides us with a conceptually much cleaner solution. Following the magnetostatics analogy, which, for the incompressional part of the fluid flow, is {\em exact}, we simply have
\be \label{v curl A}
\vec v_0 = \vec \nabla \times \vec A \; .
\ee
So, the local action that describes vortex lines, the hydrophoton, sound, and their interactions, is simply
\begin{align} \label{local action with sound}
S = & w_0 \bigg[- \sum_{n} \Gamma_n  \int dt  d\lambda     \,  \sfrac{1}{3} \epsilon^{ijk} \, X_n^i \, \di_t \, X_n^j \, \di_{\lambda} X_n^k  \\
 & +  \int d^3 x dt \, \Big[ \big( \di_i A_j  \big)^2 + \sfrac{1}{2} \dot{\vec  \psi} \, ^2-\sfrac12 c_s^2  (\vec \nabla \cdot \vec \psi)^2 \Big] \nonumber\\
  & + \sum_{n}  \Gamma_n  \int dt d\lambda\,  \di_\lambda \vec X_n \cdot \vec A\big( \vec X_n, t \big) +  \int d^3x dt \,  (\vec \nabla \times \vec A)^i  \big((\vec \nabla \times \vec A) \cdot \vec \nabla \big)\psi^i + \dots  \bigg] \; . 
  \nonumber
\end{align}
The first line contains the vortex lines' peculiar kinetic terms. The second line is the free action for the hydrophoton and sound fields. The third line contains all the interactions: of $\vec A$ with the vortex lines, and of $\vec A$ with sound. The dots denote all higher order sound interactions of Appendix \ref{expanding L}, which, given the replacement \eqref{v curl A}, involve more powers of $\vec \psi$ and $\vec A$, and are straightforward to write down explicitly.

\begin{figure}
\begin{center} 
\includegraphics[width=0.2\textwidth]{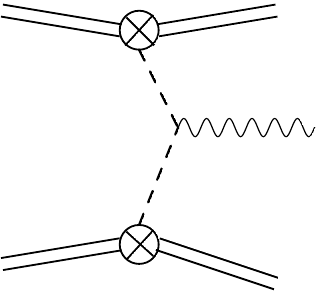}
\end{center}
\caption{\small \em Lowest order Feynman diagram contributing to the emission of sound by the scattering of two vortex rings. The dashed lines represent the hydrophoton and the wavy line the sound. }
\label{Hydrophoton_Sound}
\end{figure}

The message is clear: vortex lines interact directly with $\vec A$; sound interacts directly with $\vec A$; vortex lines and sound interact only indirectly, via $\vec A$. For processes that are amenable to a perturbative treatment, one can then use standard field theoretical/diagrammatic techniques to compute observables. Consider for instance a soft scattering process in which two vortex rings pass by each other at large impact parameter (compared to their radii). What is the sound emitted by such a ``collision''? The most relevant Feynman diagram is depicted in fig.~\ref{Hydrophoton_Sound}: the two vortex rings exchange a long-distance hydrophoton, and this internal line emits a phonon, as dictated by the $AA\psi$ vertex in the action above. 

\section{Discussion and Outlook}

We have initiated a systematic application of effective field theory techniques to the study of near incompressible hydrodynamical systems. As usual for local field theories,  questions that can be dealt with in perturbation theory are reduced, at least conceptually, to ``turning the crank'':  one can organize each observable  as an expansion in Feynman diagrams---of ever increasing complexity as one moves to higher orders in perturbation theory---each of which can be computed in standard ways starting from the action.

The small expansion parameters in our analysis have been the typical fluid-flow speed, which we took by assumption to be much smaller than that of sound, and the typical size of the regions where vorticose flow is localized, which we took to be small relative to the sound waves' wavelengths and to the distance to other vorticose regions. The smallness of the former parameter allows for a perturbative expansion in compressional modes (sound). The smallness of the latter allows for a multipole expansion.

Our insisting on the field theory language has also led us to an entirely {\em local} action for vortex-line dynamics---which is nontrivial, given that, as is well known, each line carries a long-distance $1/r$ velocity profile with it. We have `integrated in' such a velocity profile---or better, the associated vector potential (which we have dubbed the ``hydrophoton")---thus ending up with a local field theory describing vortex lines, the hydrophoton field, sound modes, and their mutual interactions, to {\em all} orders in $(v/c_s)$. In this reformulation, vortex lines interact with themselves, with one another, and with sound, by exchanging hydrophotons. We believe this to be substantial technical improvement over the standard ``vortex filament" model. Once again, questions amenable to a perturbative expansion are straightforward to analyze---and answer.

From a theoretical viewpoint, there are a number of open conceptual questions about our field-theory formulation of vortex line dynamics. First, we did not construct it bottom-up from standard low-energy effective field theory principles, nor did we derive it top-down from our general fluid action \eqref{action}---we just guessed the vortex-line action, and checked that it reproduces the correct dynamics as derived from the Euler equation.
To be responsible field theorists, we should remedy this. In particular, as we emphasized, the dynamics of vortex lines are strikingly different from those of string-like objects in empty space. Vortex lines obey first order equations of motion, which means that their local velocity is completely determined by the geometry of the lines' configuration. 
According to standard effective field theory intuition, all robust properties of the dynamics should follow from symmetry---in order not to be disrupted by renormalization, for instance. So, what is the symmetry protecting such a peculiar behavior? Compared to string-like objects in empty space, we seem to have {\em less} symmetry---Lorentz or Galileo boosts are gone, being (spontaneously) broken by the surrounding medium. This allows us to write down {\em more} Lagrangian terms, including our peculiar kinetic term with one time derivative---but the action \eqref{3D action for two vortex lines}, or equivalently \eqref{local action}, is hardly the most generic action invariant under translations and rotations. This is all the more frustrating as our original fluid action, eq.~\eqref{action}, was constructed precisely as the most generic action invariant under all the appropriate symmetries. Reparameterization invariance under $\lambda \to \tilde \lambda(\lambda, t)$---which is  symmetry of \eqref{3D action for two vortex lines} and  \eqref{local action}---will play of course a crucial role in reducing the number of allowed Lagrangian terms. We plan to address these symmetry questions in the near future.

Second, for small vortex rings we would like to go one step further, and treat them as point-like objects with multipole moments (starting with a `magnetic' dipole), and a number of degrees of freedom: their position, radius, orientation, and an infinite tower of excited states (Kelvin waves). 
Once again, given the first-order nature of the equations of motion, the dynamics of these objects are quite peculiar. For instance, from a preliminary analysis it seems to us that the orientation is {\em not} a low-energy degree of freedom: one can certainly choose any orientation as an initial condition, but then one cannot change it via low-frequency processes. In a sense, unlike for ordinary string loops in empty space, the orientation degree of freedom seems to belong in the excited Kelvin wave spectrum. Again, all the vortex ring properties should follow from systematically applying symmetry considerations  to the effective theory of point-like objects in the fluid, along the lines of \cite{GR}. 

Third, what is this hydrophoton anyway? We have a local field $\vec A$, whose equation of motion in the incompressible limit, 
\be
\nabla^2 \vec A = \vec \omega  \; ,
\ee
implies instantaneous signal propagation at arbitrarily large distances. This is clearly an artifact of the incompressible approximation, and once the speed of sound  is brought down to finite values, we expect $\vec A $ to propagate at that speed---no small disturbances can propagate faster. This sounds very much like {\em sound}. Is $\vec A$ in some sense sound itself? 

But, while we responsibly address all the conceptual points raised above, nothing prevents us from looking for relevant physical applications of our methods, which can range for instance from classical vortex ring systems in ordinary fluids like water \cite{WIrvine_DKleckner}, to quantized vortex line interactions in laboratory superfluids or in pulsars, where dense arrays of vortex lines in the neutron superfluid are expected to interact strongly with much denser arrays of magnetic flux tubes in the proton superconductor, with crucial consequences for the dynamics of the star as a whole \cite{Mal}.

%
%
%
%
%
%
%

\section*{Acknowledgments}

We would like to thank Andrej Ficnar, J.~C.~S\'eamus Davis, Walter Goldberger,  Riccardo Penco, Rafael Porto, Ira Rothstein, Mal Ruderman, and Itamar Yaakova for stimulating and enlightening conversations.
We are especially thankful to William Irvine and Slava Rychkov for extensive discussions on the dynamics of vortex lines and vortex rings.
The work of SE is supported by the NSF through a Graduate Research Fellowship. The work of AN is supported by the DOE under contract DE-FG02-11ER1141743 and by NASA under contract
NNX10AH14G.

\appendix

\section{Systematic expansion of the Lagrangian}
\label{expanding L}

Here we make more explicit the steps taken to arrive at the foundation of the work in this paper. We have attempted to make this section as complete as possible, and as a result there is some  overlap with the more rapid expansion found in Section \ref{coupling of vortex and sound}, and with refs.~\cite{DGNR, ENRW}.

We begin with the field theoretical description for a relativistic perfect fluid \cite{ENRW, DGNR, Soper}. Its action is given in eq.~\eqref{action}, which we reproduce here for completeness:
\be \label{action_appendix}
S = \int \! d^4 x \, F(b) \; , \qquad \text{with} \quad b \equiv \sqrt{\det B^{IJ}} \; ,
\ee
where $B^{IJ}=\di_\mu \phi^I \di^\mu \phi^J$ and where $F$ is a generic function. A particular functional form of $F$ corresponds to a particular equation of state. However, in order carry out the expansion of the Lagrangian in fluctuations about a background flow, there is a slightly more convenient starting point than eq.~(\ref{action_appendix}).
Manipulating $b$ we can pull out the Jacobian determinant for the equal-time $\vec x \leftrightarrow \vec \phi$ mapping, which, for nearly incompressible flows, is close to one
\footnote {This separation is also convenient for expanding around a static background, $\phi^I=x^I$, and was utilized in \cite{ENRW}.}:
\bea
b^2 & = & \det{\di_\mu \phi^I \di^\mu \phi^J} \nonumber \\
&= & \det \big( \di \phi^T \cdot \di \phi - \dot {\vec \phi} \otimes \dot {\vec \phi} \big) \nonumber \\
& = & \det \Big(  \di \phi^T \cdot \big(1 - ((\di \phi^T)^{-1} \cdot \dot {\vec \phi}) \otimes  ((\di \phi^T)^{-1} \cdot \dot {\vec \phi})  \big) \cdot  \di \phi\Big) \nonumber \\
& = & (\det \di \phi)^2 \, \det\big(1 -\vec v \otimes \vec v \big) 
\label{B} \;,
\eea
where we defined the matrix $(\di \phi)_{ij} \equiv \di_i \phi_j$, and the vector
\be
\vec v \equiv - (\di \phi^T)^{-1} \cdot \dot {\vec \phi} \; .
\ee
Notice that, by the implicit function theorem, $\vec v$ is precisely the usual fluid velocity field---hence the name:
\be
\vec v = \frac{\di \vec x (\vec \phi, t)}{\di t} \Big|_\phi \; .
\ee

The second determinant in (\ref{B}) is easy to compute by going to a basis where, locally, the $x$ axis is aligned with $\vec v$. We get
\be
b^2 = (\det \di \phi)^2 \big(1 - |\vec v|^2 ) 
\ee
and therefore we can rewrite the Lagrangian in the  form:
\be \label{expandL}
{\cal L} =  - w_0 c^2  f \big(\det \di \phi \,  \sqrt{1 - |\vec v|^2} \, \big) \; ,
\ee
where $f$ and $w_0$ are defined in eqs.~\eqref{w0}, \eqref{f(b)}.
As $\vec \phi( x,t)$ is one-to-one with $\vec x$ for all $t$, we can change coordinates to those of the comoving volume elements and we have arrived at equation (\ref{comoving action}):
\be  \label{appendix action}
S  =   - w_0 c^2 \int \! d^3 \phi dt \, \det J \,f\big( ( \det J^{-1}) \sqrt{1- v ^2/c^2} \big)  \; .
\ee
$J^i {}_j$ is the Jacobian matrix $\di x^i / \di \phi^j$, and $\vec v = \di_t {\vec x} ( \phi, t)$. 

We now expand this action in powers of $v/c_s$---which is small by assumption---and of the compressional field $\vec \psi$, which, as explained in sect.~\ref{coupling of vortex and sound}, describes the small deviations from a purely volume-preserving time evolution:
\be
\vec x( \phi,t)=\vec x_0( \phi,t)+\vec \psi( \phi,t) \; , \qquad \det \left(J_0\right)\equiv \det \left( \frac{\di x_0^i}{\di \phi^j}\right)=1 \; , 
\qquad \vec \psi = \vec \nabla_0 \Psi\; .
\ee
For $\vec v$ and  $\det J $ the expansion in $\vec \psi$ truncates at finite order (the determinant of a $3\times 3$ matrix is a cubic function of that matrix):
\begin{align}
& \vec v  = \vec v_0 + \di_t \vec  \psi \, \big |_\phi \;, \qquad \vec v_0 \equiv \di_t \vec x_0 \, \big |_\phi \\
& \det J =  1+[\nabla_0 \psi]+\sfrac{1}{2}\left([\nabla_0 \psi]^2-[\nabla_0 \psi^2] \right)+\sfrac{1}{6}\left([\nabla_0 \psi]^3-3[\nabla_0 \psi][\nabla_0 \psi^2]+2[\nabla_0 \psi^3] \right) \; ,
\end{align}
where $[...]$ means the trace, $\nabla_0 \psi^n$ means the $n$-th power of the matrix $\di \psi^i / \di x_0^j$, and we have used the identity
\be
J^i {}_j \equiv \frac{\di x^i}{\di \phi^j } = \Big(\delta^i_k + \frac{\di\psi^i}{\di \phi^l} \frac{\di \phi^l}{\di x_0^k}    \Big) \, {J_0}^{k} {}_j 
= \Big(\delta^i_k + \frac{\di\psi^i}{\di x_0^k}    \Big) \,  {J_0}^{k} {}_j \; .
\ee
Of course, for the inverse determinant and  the square root entering the action \eqref{appendix action} the expansion does not truncate, and goes on to all orders. Up to cubic order in $\vec \psi$ we get
\begin{align}
\label{expanded L in comoving coords}
S = 
w_0\int d^3\phi \,dt\; \Big\{ 
& + \sfrac12 {v^2}+\sfrac{(c^2-c_s^2)v^4}{8c^4}  \\
& +\sfrac{1}{2}\left((\di_t \psi)^2-c_s^2[\nabla_0 \psi]^2 \right) \nonumber \\
&+ v_i \, \di_t \psi^i -\sfrac12\sfrac{c_s^2}{c^2} v^2 \, [\nabla_0 \psi] \nonumber \\
&  +\frac{1}{2c^4}(c^2-c_s^2)v^2 v_i \, \di_t \psi^i+\frac{1}{8c^2}f_3 \,  v^4 \, [\nabla_0 \psi] \nonumber\\
& -\frac{c_s^2}{c^2} v_i \di_t \psi^i [\nabla_0 \psi]+\frac{1}{4c^2}\left( c_s^2v^2([\nabla_0 \psi]^2+[\nabla_0 \psi^2])+c^2 \, f_3 \, v^2[\nabla_0 \psi]^2 \right) \nonumber \\
& + \frac{1}{6c^2}\left(-3c_s^2 [\nabla_0 \psi]\left( (\di_t \psi)^2-c^2[\nabla_0 \psi^2] \right)+c^4 \, f_3\, [\nabla_0 \psi]^3 \right)   + \dots \Big\}\;. \nonumber
\end{align}
We have used $f'(1) =1$, $f''(1) = c_s^2/c^2 $, $f_3 \equiv f'''(1)$, and for notational simplicity we have dropped the subscript zero in $v_0$, and so, from now on, $v \equiv v_0$ is the underlying incompressional velocity field. Even though $f_3$ is a free parameter, as pointed out in \cite{ENRW} one typically expects
\be
f_3 \sim c_s^2/c^2 \; .
\ee
We have dropped total derivative terms, 
most notably the term  $-w_0 c^2 \, f(1) \,  \det J$, since $\det J$ itself is a total derivative---schematically: 
\be
\det J = \det \di_\phi x  \propto \epsilon \epsilon \cdot \di_\phi x \, \di_\phi x \, \di_\phi x =  \di_\phi \cdot \big( \epsilon \epsilon \cdot  x \, \di_\phi x \, \di_\phi x \big) \; .
\ee
We have also dropped terms that are total derivatives {\em with respect to $\vec x_0$}, even though the integral is in $d^3 \phi$, for reasons that will become clear in a moment.
Notice that for different powers of $\psi$, we have stopped the expansion at different orders in $v_0$. The reason is that the two expansion parameters are not entirely independent, as we will explain below.
Notice also that the first line is just the action for the incompressible fluid flow, in the absence of compressional perturbations. To be used in this sense, it should be supplemented by a volume-preserving constraint for $\vec x_0(\vec \phi, t)$, as explained in the main text. However that is not our goal, since we are interested in solving for the dynamics of $\vec \psi$ in the presence of a {\em given} background incompressible fluid flow $\vec v_0$. We will therefore discard the first line for what follows.

We are not finished yet.  We want to now go into the $\vec x_0$ coordinate system.
The reason is that in an experiment it is much more convenient to parameterize  a velocity field in terms of its dependence on the physical $\vec x$ coordinates  rather than on the comoving ones. Since the $\vec x$ and $\vec x_0$ coordinate systems coincide at lowest order in $\vec \psi$,  using $\vec x_0$ will suffice for lowest order computations.
This is an easy transformation for most things in the action: The Jacobian $J_0$ has unit determinant, and all the spacial partial derivatives are with respect to $\vec{x}_0$ already (this is the reason why above we were allowed to neglect total $x_0$-derivatives). The non-trivial pieces are the partial {\em time} derivatives as they are taken at constant $\phi$ values. We can write the time derivatives in a more compatible form:
\be
\di_t \psi^i(x_0(\phi,t),t)|_{\vec \phi}=\di_t \psi^i(x_0,t)|_{x_0}+v_0^j \frac{\di}{\di x_0^j}\psi^i(x_0,t) \;.
\ee

So, changing coordinates from $\vec{\phi}$ to $\vec{x_0}$ and inserting the expanded time derivatives in the action above we arrive at an action which can be organized in the following way:
\be
S=S_{\psi^2}+S_{\psi^3}+ \dots  +S_{\psi \; v^n}+S_{\psi^2 \; v^n}+ \dots  \;,
\ee
where the $n$'s above mean `positive powers of $v$' (1, 2, 3, etc.). Dropping for notational convenience all the subscript $0$'s, and denoting partial time-derivatives by overdots, we have explicitly:
\begin{align} \label{action_in_v/c_s_start}
S_{\psi^2}&= w_0 \int d^3xdt \; \frac{1}{2} \big( \dot {\vec \psi} \, ^2-c_s^2[\di \psi]^2\big)\\
S_{\psi^3}&= w_0 \int d^3xdt \; \Big\{\frac{c_s^2}{2}[\di\psi][(\di \psi)^2]+\frac{c^2 f_3}{6}[\di \psi]^3-\frac{c_s^2}{2c^2} \dot {\vec \psi} \, ^2 [\di \psi] \Big\}\\
& \vdots  \nonumber \\
S_{\psi \; v^n}&= w_0 \int d^3xdt \; \Big\{ \vec v \cdot  \dot {\vec \psi} +v_i(v\cdot\nabla)\psi^i -\frac{c_s^2 }{2c^2}\,  v^2 [\di \psi] \nonumber \\
&+\frac{(c^2-c_s^2)}{2c^4} \big(v^2 \vec v\cdot  \dot {\vec \psi} + v^2 v_i (v \cdot \nabla)\psi^i \big) +\frac{f_3 }{8c^2} \, v^4[\di \psi]+ \cdots \Big\}\\
S_{\psi^2 \; v^n}&= w_0 \int d^3xdt \; \Big\{\dot \psi_i(v\cdot \nabla)\psi^i-\frac{c_s^2}{c^2} \big(\vec v \cdot \dot{\vec \psi } \,  \big) [\di \psi]+\frac{1}{2}\left( (v\cdot \nabla)\psi^i\right)^2 \nonumber\\
&  -\frac{c_s^2}{c^2}v_i(v\cdot \nabla)\psi^i [\di \psi]+\left(\frac{c_s^2}{4c^2}+\frac{f_3}{4} \right)v^2[\di\psi]^2+\frac{c_s^2}{4c^2}v^2 [(\di \psi)^2 ] \nonumber \\
& \frac{1}{2c^4}\left(c^2-c_s^2 \right)\big(\vec v \cdot \dot{\vec \psi } \,  \big) ^2+\frac{1}{4c^4}\left(c^2-c_s^2 \right)v^2\dot {\vec \psi} \, ^2+ \cdots \Big\}
\label{action_in_v/c_s_end} \\
& \vdots  \nonumber
\end{align}
While all of the above looks like a total mess (and of course there is an infinite tower of terms), it is both easy to generate and easy to interpret perturbatively. In particular: 
\begin{itemize}
\item
$S_{\psi^2}$ describes the free propagation of compressional modes (sound waves); $c_s$ is indeed their propagation speed. The associated (Feynman) propagator is
\be
\langle T (\psi^i \psi^j) \rangle = \frac{\hat p^i \hat p^j}{w_0} \frac{i}{\omega^2 - c_s^2 \vec p \, ^2 + i \epsilon} \; .
\ee
\item
$S_{\psi^3}$ describes the sound waves' trilinear self-interactions; in a Feynman diagram these will correspond to a vertex with three lines attached.
\item
$S_{\psi \; v^n}$ corresponds to `tadpole' diagrams for $\psi$---diagrams with a single $\psi$ line attached to the external source $\vec v_0$: In the presence of a non-trivial background incompressional velocity field, $\psi = 0 $ is not a consistent solution. Terms in the action that are linear in $\psi$ describe, at lowest order in $\psi$, how a non-trivial $\psi$ field is generated by such a background velocity field. 
\item
$S_{\psi^2 \; v^n}$ corresponds to interaction vertices where two $\psi$ lines attach to the external source $\vec v_0$, describing for instance the scattering of sound waves by vorticose motion, as in sect.~\ref{Scattering}.
\item And so on.
\end{itemize}
As discussed throughout the text, note that the terms with the additional $c_s^2/c^2$ factors scale like many of the others in terms of powers of $v/c_s$, derivatives, and so forth. These terms are the relativistic corrections, and will be necessary to correctly describe ultra-relativistic fluids.

By comparing $S_{\psi^2}$ and $S_{\psi \; v^n}$ we see that, for small $v/c_s$,  the $\psi$ field generated by the background fluid flow scales like $\psi \sim v^2$ (we are implicitly assuming that $\dot v$ scales like $v^2$, keeping the typical length scales fixed.)
According to this power-counting scheme, the terms explicitly displayed in \eqref{action_in_v/c_s_start}--\eqref{action_in_v/c_s_end}
are all the terms in the action up to order $v^6$. Things become more complicated when not all of the $\psi$'s appearing
in a diagram are ``generated" by the background $v_0$, like for instance in a scattering process. Also, in general there are important power-counting differences between highly off-shell internal lines with $\omega \ll c_s p$, and on-shell external lines with $\omega = c_s p$. These subtleties for  power counting in classical perturbation theory are addressed systematically for general relativity in ref.~\cite{GR}. It would be interesting to derive analogous power counting rules for our case. We leave this for future work, since for our lowest-order computations in this paper a quick case-by-case analysis suffices.

\section{Scattering cross section and emission rate from the amplitude}
\label{Scattering cross section from amplitude}
For the benefit of the reader we derive the scattering cross section formula for one-to-one scattering. Following Srednicki's \cite{Srednicki} quick construction---which avoids the use of wave packets---we start with the probability of scattering:
\be
P=\frac{\big| \left<f\big| i\right> \big|^2}{\left< f\big| f\right>\left< i\big| i\right>} \equiv \frac{\big| \mathcal{M} \big|^2}{\left< f\big| f\right>\left< i\big| i\right>} \;.
\ee
Note that our $\mathcal M$ here is different then the usual one because we have not yet removed any delta-functions. Assume that we are performing our experiment in a large box of volume $V$ and over a time $T$. The norm of a single particle state is given by (with the so-called relativistic normalization, which is in fact still convenient for non-relativistic systems \cite{ENRW})
\be
\left< k\big| k\right> =  2 E \, (2\pi)^3\delta^3(0) = 2 E V \; .
\ee

For our $1\to 1$ scattering off external sources of sect.~\ref{Scattering},
let's take the incoming momentum and energy to be $p_1$ and $E_1$, and the outgoing ones to be $p_2$ and $E_2$. Additionally, we want to sum over final momentum states
We then have
\bea
P=\sum_{\vec p_2} \frac{\big| \mathcal{M}(p_1,p_2)\big|^2}{2 E_1 2E_2 V^2} \; , 
\eea
which is in the continuous limit becomes
\bea
P &\rightarrow&\frac{V}{(2\pi)^3} \int d^3 p_2 \frac{\big| \mathcal{M}(p_1,p_2)\big|^2}{2 E_1 2E_2 V^2} \; .
\eea

To get the total cross section we need to divide by the incoming flux (which for an incoming particle moving at speed $c_s$ is simply $(c_s/V)$) and divide by the total time $T$. Finally, we arrive at the general expression
\bea
\label{general one-to-one scattering}
d\sigma_{(1\rightarrow1)}=\frac{1}{c_s}  \frac{1}{2E_1}  \frac{d^3p_2}{(2\pi)^3 2E_2}   \frac{\big| \mathcal{M}(p_1,p_2)\big|^2}{T}
\eea
${\cal M}$ can be computed following the standard {\em relativistic}  Feynman rules \cite{ENRW}, with the caveat that whenever an external source $J(x)$ appears in a Lagrangian term, its (four-dimensional) Fourier transform should appear as a factor in ${\cal M}$. As a consistency check, notice that for Lagrangian terms with {\em no} external sources, we formally have $J(x) = 1$, whose Fourier transform is the usual $(2 \pi)^4 \, \delta(\mbox{energy}) \delta^3(\mbox{momentum})$.

Now, if our vorticose source is time-independent, then we have an energy-conserving delta function in our $\mathcal M$, such that
\bea
\mathcal M &\rightarrow& \mathcal M' (2\pi) \delta(E_1-E_2)\\
\Rightarrow \quad \big|\mathcal M \big|^2 &\rightarrow& \big| \mathcal M' \big|^2 (2\pi) \delta(E_1-E_2) (2\pi) \delta(0)\\
&=&\big| \mathcal M' \big|^2 (2\pi) \delta(E_1-E_2)\,  T
\eea
and thus, in this time-independent limit we have that 
\bea
d\sigma_{(1\rightarrow1)}(\text{time independent source})=\frac{1}{c_s}  \frac{1}{2E_1}  \frac{d^3p_2}{(2\pi)^3 2E_2} \big| \mathcal{M'}(p_1,p_2)\big|^2 (2\pi) \delta(E_1-E_2)
\eea
which, as a good check, matches Peskin and Schroeder's Rutherford scattering problem in their fourth chapter \cite{PeskinSchroeder}.

Similar considerations apply to the emission process discussed in sect.~\ref{Sound emitted by a vorticose source}, and yield eq.~\eqref{decayrate}.

\section{Scattering sound waves off sources: Checks}

\subsection{Discrepancy with previous results}
\label{LundRojas matching}

The results found in \cite{LundRojas} were developed in the context of classical scattering theory with the same set of assumptions as ours: in particular, low Mach number of the source flow, low intensity and high frequency relative to the typical source frequency of the incoming sound waves. Moreover, the computations of ref.~\cite{LundRojas} are appropriate for  non-relativistic fluids only, so for the purposes of this comparison we will set $c_s/c \to 0$ in our results.

The first step in the comparison is to note that the vorticity field, defined as $\vec w = \nabla \times \vec v$, satisfies
\be
\left(\hat{p}_1\times \hat{p}_2 \right) \cdot \tilde{w}(\Delta p^{\mu})=i\left[ (\hat{p}_1 \cdot \vec{\Delta p})(\hat{p}_2 \cdot \tilde{v})-(\hat{p}_2 \cdot \vec{\Delta p})(\hat{p}_1 \cdot \tilde{v})\right]
\ee
where $\tilde{w}(\Delta p^{\mu})$, $\tilde{v}(\Delta p^{\mu})$ are  four-dimensional Fourier transforms, and for brevity we are using a relativistic notation, $\Delta p^\mu \equiv (\Delta \omega, \Delta \vec p \, )$). Defining the scattering angle through the usual $\hat p_1 \cdot \hat p_2=\cos \theta$, we can write
\be
\label{vorticity and velocity}
\left(\hat{p}_1\times \hat{p}_2 \right) \cdot \tilde{w}(\Delta p^{\mu})=i\left(1-\cos \theta \right) p_1 \left[(\hat p_1 +\hat p_2)\cdot \tilde v \right]+\mathcal{O}\left(\frac{\Delta \omega}{c_s p_1}\right) \; .
\ee
Here $\Delta \omega/c_s p_1$ is small---see sect.~\ref{Scattering}---and so we will neglect it in the following.

Now, using (\ref{vorticity and velocity}) we can re-express (\ref{final result--sound scattering off source}) as
\bea
\frac{d\sigma}{d\Omega \, d(\Delta \omega)} \simeq \frac{1}{4 c_s^4} \cdot\frac{\omega^2}{(2\pi)^3 } \cdot\frac{1}{T} \cdot \frac{\cos^2 \theta}{(1-\cos \theta)^2}  \left| (\hat p_1 \times \hat p_2 )\cdot \tilde{w}(\Delta p^\mu)\right|^2+\mathcal{O}\left(\frac{\Delta \omega}{c_s p_1}\right)
\eea
Using our notation for the Fourier transform (which is different than that of \cite{LundRojas}) we can express
\bea
\tilde{w}_k (\Delta p^\mu) \tilde{w}_l^*(\Delta p^\mu) &=&\int d^4x \,  d^4y\; e^{-i \Delta p^\mu x_\mu +i \Delta p^\mu y_\mu} \, w_l (y^\mu)w_k(x^\mu)\\
&=&\int d^3 R \; d \Delta \; e^{i \left(\Delta \omega \cdot \Delta-\vec{\Delta p} \cdot \vec R\right)} \\
&&\times  \int d^3 x\; \int_{-\infty}^\infty d\tau \; w_l(\vec R+ \vec x, \tau +\Delta/2) w_k( \vec x, \tau-\Delta/2) \nonumber
\eea
where we have shifted the integral variables appropriately.
Using the notation of \cite{LundRojas} we can write our final scattering cross section as
\bea
\label{our result in LR notation}
\frac{d\sigma}{d\Omega_2 d\nu}= \frac{ (2\pi) \nu^2}{4c^4 } \cdot \frac{\cos^2 \theta}{(1-\cos \theta)^2} \, A_k A_l \, \tilde{S}_{kl}(\vec q, \nu-\nu_0)
\eea
where $\nu = \omega_2$, $\nu_0 = \omega_1$, $\vec q=\vec{\Delta p}$, $d\Omega_2 = d \Omega$, $A_k=(\hat p_1 \times \hat p_2)_k$,  
\be
\tilde{S}_{kl}(\vec q, \nu-\nu_0)=\frac{1}{(2\pi)^4}\int d^3r \; d \Delta \; e^{i \left(\Delta E\cdot \Delta-\vec{\Delta p} \cdot \vec R\right)} \int d^3x \left< w_k(\vec R+ \vec x,\Delta/2)w_l(\vec x,-\Delta/2)\right> \; ,
\ee
and where 
\be
\left< f(\Delta/2)f(-\Delta/2)\right> \equiv \frac{1}{T}\int_{-T/2}^{T/2}d\tau f(\tau+\Delta/2)f(\tau-\Delta/2) \; .
\ee
When we compare (\ref{our result in LR notation}) to that of equation (19) in \cite{LundRojas} we see that there is a factor of $2$ discrepancy.



\subsection{Agreement with  optical theorem}
\label{optical theorem check}

The optical theorem as expressed in quantum field theory is a consequence of the unitarity of the {\em S}-matrix. In our case, it relates the imaginary part of the one-to-one amplitude in the forward scattering limit,  to the sum over the squares of the one-to-anything amplitudes\footnote{We use the conventions given in \cite{Srednicki} and \cite{PeskinSchroeder}, which should be noted are different than those given in \cite{Weinberg}.}:
\be
2 \text{ Im } \mathcal{M}(i \rightarrow i) =\sum_f \int d\Pi_f \; \left| \mathcal{M} (i \rightarrow f) \right|^2 \; .
\ee 
It is a statement about full amplitudes, but perturbatively it can be expressed as a relation between particular graphs. To lowest order (in the coupling) the relationship is simply given by:
\be
\label{Pert Opt Theorem}
{2 \text{ Im }} \left( \begin{array}{c}\includegraphics[height=1.3cm]{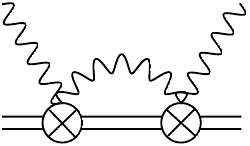} \end{array} \right)={\int \frac{d^3q}{(2\pi)^3 2 \omega}} \left| \begin{array}{c}\includegraphics[height=1.3cm]{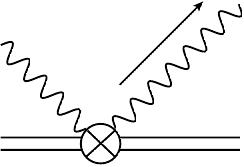} \end{array}\right|^{{2}} \;.
\ee
We can rewrite the right hand side of the above equation as the one-to-one scattering cross-section using (\ref{general one-to-one scattering}). Written this way, we have
\be
\label{Pert Opt Theorem Mod}
2 \text{ Im } \mathcal{M} (\vec p_1,\vec p_1) =2 \omega_1 c_s T \, \sigma_{(1 \to 1)}  \; .
\ee
Eq.~(\ref{final result--sound scattering off source}) provides us with the right hand side of this equation. 
We need to calculate the left hand side.

The imaginary part of the amplitude comes from the $ i \epsilon$ prescription in the internal propagator of sound modes. 
Using generating functional methods to ensure that all symmetry factors are correct, we have
\bea
\left<0 \left| T \psi^i (x_1) \psi^j(x_2) \right |0 \right>_{v}&=& \frac{1}{i}\frac{\delta}{\delta J^i(x_1)} \cdot \frac{1}{i}\frac{\delta}{\delta J^j(x_2)} \cdot \frac{1}{2!} \left( i \int d^4y\; \mathcal{L}_{\psi^2  v}\left(\frac{1}{i}\frac{\delta}{\delta \vec J(y)} \right)\right)^2  \nonumber \\
&& \times \frac{1}{3!} \cdot \left( \frac{i}{2} \int d^4x d^4x' J^k(x) \Delta^{kl}(x-x')J^l(x')\; \right)^3\;.
\eea
Upon taking all the functional derivatives it is easy to see that one is left with four terms. Inserting this correlation function into the LSZ formula the amplitude is
\bea
i \mathcal{M}(\vec p_1 \rightarrow \vec p_2) &=& i \int \frac{d \omega_q}{2\pi} \frac{d^3 q}{(2 \pi)^3} \,  \frac{1}{-\omega_q^2+c_s^2 \vec q \,^2- i \epsilon} \int_{y_1 y_2} e^{iy_1 (p_1-q)} e^{iy_2(q-p_2)} \nonumber \\
&&\times \left[\omega_1 \left(\vec v (y_1) \cdot \vec q\right)+ \omega_q\left(\vec v (y_1) \cdot \vec p_1\right) \right] (\hat q \cdot \hat p_1) \nonumber \\
&& \times \left[\omega_2 \left(\vec v (y_2) \cdot \vec q\right)+\omega_q\left(\vec v (y_2) \cdot \vec p_2\right) \right] (\hat q \cdot \hat p_2) \; ,
\eea
where we have suppressed relativistic corrections for algebraic simplicity; their inclusion is straightforward. 

Taking the forward scattering limit, $p_2\rightarrow p_1$, and performing the $y$ integrals which yield Fourier transforms of the velocity fields we arrive at
\be
i \mathcal{M}(\vec p_1 \rightarrow \vec p_1) = i \int \frac{d \omega_q}{2\pi} \frac{d^3 q}{(2 \pi)^3} \, \frac{1}{-\omega_q^2+c_s^2 \vec q^2- i \epsilon} \tilde{v}^i(q-p_1) \tilde{v}^j(p_1-q)V^i V^j ,
\ee
where $V^i=(\hat q \cdot \hat p_1) \left[ \omega_1 q^i+ \omega_q p_1^i\right]$. In this form it is easy to see that the only contribution to the $\text{Im } \mathcal{M}$ comes from the $i \epsilon$, we have
\bea
\label{Im M with 2 deltas}
\text{ Im } \mathcal{M}(p_1,p_1) & =&  \int \frac{d \omega_q}{2\pi} \frac{d^3 q}{(2 \pi)^3} \frac{\pi}{2c_s |\vec q|} \big[\delta(\omega_q -c_s |\vec q| )+\delta(\omega_q +c_s |\vec q| ) \big] \nonumber \\
& \times & \tilde{v}^i(p_1-q) \tilde{v}^j(q-p_1)V^iV^j \;,
\eea
where we have isolated the imaginary part with the standard formula
\be
\frac{1}{x\pm i \epsilon}=\mathcal{P} \frac1x \mp  i \pi \delta(x) \; ,
\ee
where $\mathcal{P}$ is the principal value.

Only the first delta-function in (\ref{Im M with 2 deltas}) is going to overlap with the support of $\tilde{v}^i(p_1-q)$. This is because $\tilde{v}^i(p_1-q)$ offers a narrow width around zero frequency relative to the incoming frequency (as the time scale of the velocity flow is much longer than the incoming frequency). For instance, in the static limit $\tilde{v}^i(p_1-q) \propto \delta(\omega_1- \omega_q)\tilde{v}^i(\vec p_1-\vec q)$. For the same reason, as discussed in Section \ref{Scattering}, to lowest order in the energy transfer we can take $\omega_1=\omega_2=\omega$.
Evaluating this final delta function, relabeling our variables of integration and performing some trivial algebra we have:
\be
2 \text{ Im } \mathcal{M}(p_1,p_1)=(2 \omega c_s T) \int \frac{d \Omega \, d( \Delta \omega)}{(2 \pi)^3} \frac{\omega^4}{4c_s^6 T}(\hat{p}_1 \cdot \hat{p}_2)^2 \left|\tilde{v}^i(\Delta \omega, \vec{\Delta p}) \cdot \left(\hat{p}_1^i+\hat{p}_2^i\right) \right|^2
\ee
Inserting this into the left-hand side of (\ref{Pert Opt Theorem Mod}) and comparing with our scatting cross section (\ref{final result--sound scattering off source}) calculated in the bulk of the paper  we see that indeed the optical theorem is satisfied.

\end{document}